\documentclass[]{aa}  

\usepackage{graphicx}
\usepackage{txfonts}
\begin{document}

   \title{An ALMA study of the Orion Integral Filament:\\ I. Evidence for narrow fibers in a massive cloud}


   \author{A. Hacar \inst{1}
          \and
          M. Tafalla\inst{2} 
          \and 
          J. Forbrich \inst{3,4}
          \and
          J. Alves \inst{5} 
          \and
          S. Meingast \inst{5} 
          \and
          J. Grossschedl \inst{5} 
          \and
          P. S. Teixeira \inst{5,6}
          }
          \institute{
         	   Leiden Observatory, Leiden University, P.O. Box 9513, 2300-RA Leiden, The Netherlands\\
         	   \email{hacar@strw.leidenuniv.nl}
         \and
             Observatorio Astronomico Nacional (IGN), C/ Alfonso XII, 3, E-28014, Madrid, Spain
         \and 
         	 Centre for Astrophysics Research, University of Hertfordshire, College Lane, Hatfield, AL10 9AB, UK
         \and 
         	 Harvard-Smithsonian Center for Astrophysics, 60 Garden St, Cambridge, MA 02138, USA
         \and 
             University of Vienna, T\"urkenschanzstrasse 17, A-1180 Vienna, Austria
	    \and
	    	Scottish Universities Physics Alliance (SUPA), School of Physics and Astronomy, University of St. Andrews, North Haugh, St. Andrews, Fife KY16 9SS, UK }

   \date{XXXX}

 
  \abstract{}
{
  Are all filaments bundles of fibers? 
  To address this question, we have investigated the gas organization within the paradigmatic Integral Shape Filament (ISF) in Orion.
}
{
  We combined two new ALMA Cycle~3 mosaics with previous IRAM~30m observations to produce a high-dynamic range N$_2$H$^+$ (1-0) emission map of the ISF tracing its high-density material and velocity structure down to scales of 0.009~pc (or $\sim$~2000~AU).
}
{
  From the analysis of the gas kinematics, we identify a total of 55 dense fibers in the central region of the ISF. 
  Independently of their location in the cloud, these fibers are characterized by transonic internal motions, lengths of $\sim$~0.15~pc, and masses per-unit-length close to those expected in hydrostatic equilibrium. The ISF fibers are spatially organized forming a dense bundle with multiple hub-like associations likely shaped by the local gravitational potential. Within this complex network, the ISF fibers show a compact radial emission profile with a median FWHM of 0.035~pc systematically narrower than the previously proposed universal 0.1~pc filament width. 
}
{
  Our ALMA observations reveal complex bundles of fibers in
  the ISF, suggesting strong similarities between the internal substructure of this massive filament and previously studied lower-mass objects.
  The fibers show identical dynamic properties in both low- and high-mass
  regions, and their widespread detection in nearby clouds suggests a
  preferred organizational mechanism of gas in which the
  physical fiber dimensions (width and length) are self-regulated
  depending on their intrinsic gas density. Combining these results with
  previous works in Musca, Taurus, and Perseus, we identify a
  systematic increase of the surface density of fibers as a function of
  the total mass per-unit-length in filamentary clouds. Based on this
  empirical correlation, we propose a unified star-formation scenario
  where the observed differences between low- and high-mass clouds, and
  the origin of clusters, emerge naturally from the initial
  concentration of fibers.
}


   \maketitle
%

\section{Introduction}\label{sec:intro}

Investigating the internal structure of massive filaments is of crucial importance for the description of the star formation process in the Milky Way. 
As indicated by recent galactic plane surveys, 
high-mass stars and massive clusters are typically formed within filaments with total masses per-unit-length (M$_{\mathrm{lin}}$) between $\sim$~100-4000 M$_\sun$~pc$^{-1}$ \citep[e.g., ][]{PER09,MOL10,SCH14,LI16}. Typically located at kpc distances, most of these massive filaments are found in hub-like associations \citep{MYE09} forming dense ridges of gas \citep{GAL10,SCHN10,HEN10,HIL11,HEN12,PER13}. Morphological and dynamical arguments suggest a direct link between these massive filaments and their better characterized low-mass counterparts with M$_{lin}\lesssim$~100~M$_\sun$~pc$^{-1}$, regularly identified in the solar neighbourhood \citep{AND10,ARZ11,HAC11,HAC13,PAL13,HAC16}.  So far, however, the detailed comparison between these two filamentary regimes has been hampered by the resolution and sensitivity of current far-infrared (FIR) and (sub-)millimeter observations. As result, the connection between low- and high-mass filaments remains controversial 
\citep[e.g., see][]{AND14,MOT17}.

Recent molecular observations have revealed the intrinsic substructure of low-mass filaments in nearby clouds.
\citet{HAC13} demonstrated that the apparently monolithic B213-L1495 filament  \citep[M$_{lin}\sim$50 M$_\sun$~pc$^{-1}$;][]{BAR27,HAR02,PAL13} is actually a bundle of small-scale fibers.
These fibers are characterized by their continuity in space and velocity, transonic internal velocity dispersions along their typical length of $\sim$~0.5~pc, and individual M$_{\mathrm{lin}}$ consistent with hydrostatic equilibrium.
After this discovery, analogous fibers have been systematically reported in low-mass filaments (M$_{lin}\sim$~20-50 M$_\sun$~pc$^{-1}$) like \object{IC5146} \citep{ARZ13}, \object{Musca} \citep{HAC16}, and \object{TMC-1} \citep{FER16}.
Compact networks of fibers have been also identified in the NGC1333 ridge \citep[M$_{lin}\sim$~200 M$_\sun$~pc$^{-1}$;][]{HAC17b}. 
In all cases, the fibers harbour most of the cores within these regions, regulating the initial conditions for their gravitational collapse \citep[see also][]{HAC11,TAF15}. Dominating the gas substructure in isolated and clustered environments, 
fibers appear to play a fundamental role in both low- and intermediate mass filaments.

In analogy to low-mass filaments,
high-mass filaments like {\it \object{Nessie}} \citep[M$_{lin}\sim$~525 M$_\sun$~pc$^{-1}$;][]{JAC10}, \object{G11.1} \citep[M$_{lin}\sim$~600 M$_\sun$~pc$^{-1}$;][]{KAI13}, and \object{NGC6334} \citep[M$_{lin}\sim$~1000 M$_\sun$~pc$^{-1}$;][]{AND16} exhibit an increasing level of substructure at sub-parsec scales.
Additional observational evidence indicates the existence of fibers in some of these massive environments.
 Complex line profiles, containing multiple narrow velocity components, are commonly reported towards Infrared Dark Clouds (IRDC) like \object{G035} \citep[M$_{lin}\sim$~100 M$_\sun$~pc$^{-1}$;][]{HEN14}, \object{G14.225} \citep[M$_{lin}\sim$~200 M$_\sun$~pc$^{-1}$;][]{BUS13}, or \object{IRDC 18223} \citep[M$_{lin}\sim$~1000 M$_\sun$~pc$^{-1}$;][]{BEU15}.  These velocity components appear to be organized in elongated, sub-parsec scale, fiber-like threads partially resolved in recent interferometric observations \citep{HEN16,HEN17}.  
In the light of the above, can fibers also explain the internal structure of these massive clouds?

In this work (Paper I) we investigate the dense gas substructure of the paradigmatic Orion Integral Shape Filament (ISF) \citep{BAL87} combining a new set of ALMA Cycle-3 with IRAM~30m observations. 
Along its more than 7~pc of length, the ISF describes a dense ridge of gas of $\lesssim$~0.2~pc width, running approximately parallel to the declination axis at the northern end of the Orion A cloud \citep{JOH99}.
The ISF is the most massive filament among the Gould Belt clouds \citep[M$_{lin}\sim$~500 M$_\sun$~pc$^{-1}$;][]{BAL87} and the only one containing a high-mass cluster, namely, the Orion Nebula Cluster \citep[ONC;][]{ODE08}. Due to its proximity \citep[D=414~pc][]{MEN07}, the ISF is one of the best studied massive filaments and is usually employed as benchmark for clustered star-formation theories \citep[see][ and references therein]{BAL08,MUN08}.

At large scales, the gas content of the ISF has been systematically surveyed using single-dish observations of multiple CO isotopologues \citep{BAL87,DUT91,SHI11,BER14,BUC12,SHI14}, dense tracers \citep{IKE07,TAT08,HAC17a,FRI17,KAU17}, and recombination lines \citep{GOI15}. Its mass distribution has been also investigated in the continuum at both (sub-)millimeter \citep{CHI97,JOH99,SAL15} and FIR wavelengths \citep{LOM14,STU15} showing a series of clumps regularly spaced at scales of $\sim$~1~pc \citep{DUT91}, typically referred to as the OMC 1-4 clouds \citep[see also][]{PET08}. Higher resolution studies reveal a rich substructure of small-scale sub-filaments \citep{MAR90,ROD92,WIS98,LI13,HAC17a} and condensations \citep{MEZ90,CHI97} extending along the main axis of the ISF. A census of its embedded stellar population at IR \citep{MEG12,STU13,FUR16}, X-ray \citep{GET05,RIV13}, centimeter  \citep{KOU14,FOR16}, and millimeter \citep{TAK13,TEI16,KAI17,PAL17} wavelengths indicate that most of the current star formation within the ISF is concentrated towards both OMC-1 and OMC-2/3 clouds \citep{PET08}.
Focused on these two active subregions, our new N$_2$H$^+$ (1-0) ALMA observations (Sect.~\ref{sec:observations}) aim to explore the existence of fibers within this massive filament (Sect.~\ref{sec:densegas}), as well as their possible connection with the formation of massive stars and clusters (Sect.~\ref{sec:discussion}).


\section{ALMA Cycle-3 observations}\label{sec:observations}
\begin{figure*}
	\centering
	\includegraphics[width=\textwidth]{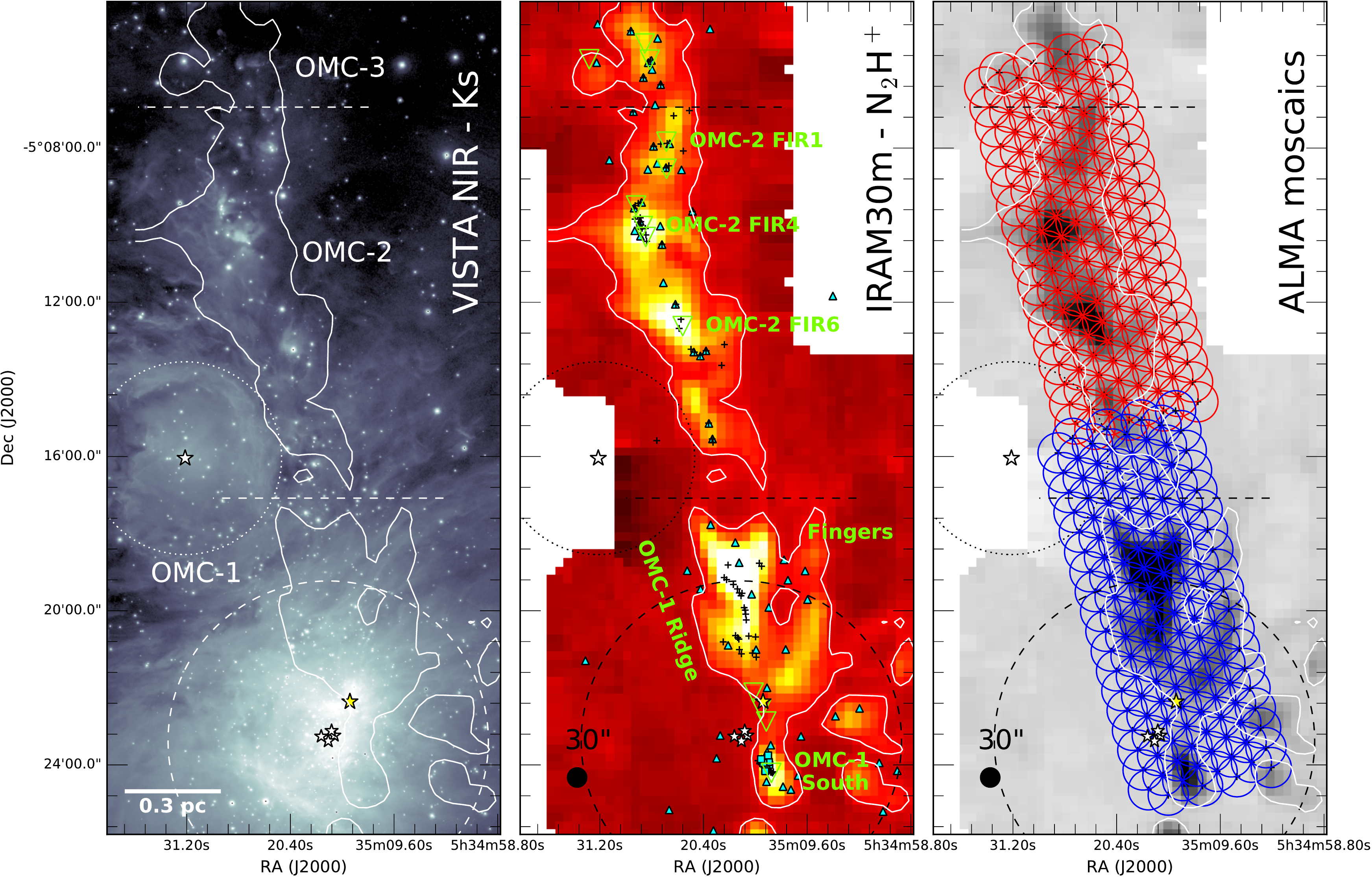}
	\caption{Description of our ALMA Cycle-3 observations along the ISF: {\bf(Left)} VISTA NIR (Ks band) emission \citep{MEI16}; {\bf(Centre)} IRAM~30m (single-dish) N$_2$H$^+$ (1-0) integrated emission \citep{HAC17a}; {\bf(Right)} 12m-array footprints of the two ALMA Cycle-3 mosaics (blue and red) presented in this work. The position of the Trapezium (white stars) and NU Ori stars (white isolated star), the Orion BN source (yellow star), and the innermost 0.5~pc radius of the ONC (dashed circle) are indicated in all panels. All figures include the contour enclosing those regions with integrated emission I(N$_2$H$^+$)~$\ge$~20 mJy~beam$^{-1}$ according to the single-dish observations. The position of all the previously identified Spitzer protostars \citep[blue triangles;][]{MEG12,STU13,FUR16}, mm-continuum peaks \citep[green triangles;][]{MEZ90,CHI97}, SMA \citep{TEI16} plus ALMA continuum sources \citep{KAI17,PAL17} (black crosses), and embedded X-ray objects \citep[blue squares;][]{RIV13} are indicated in the central panel. The most relevent regions are also labelled in both VISTA and single-dish images. For reference, a scale bar denotes the angular size of a 0.3~pc region at the distance of the ONC \citep[414~pc;][]{MEN07}. 
	}
	\label{fig:ISF_VISTA}
\end{figure*}


We mapped the central region of the ISF between December 26th, 2015 and January 2nd, 2016 using ALMA Cycle-3 observations (ID: 2015.1.00669.S; PI: A. Hacar)\footnote{
	The data products of this work are available via CDS (link).
	This work is part of the ORION-4D project (PI: A. Hacar). See more information in https://sites.google.com/site/orion4dproject .}.
As shown in Fig.~\ref{fig:ISF_VISTA}, we combined two 148-pointing ALMA mosaics, of $\sim 240"\times600"$ each, following the main axis of this cloud traced by previous single-dish observations \citep{HAC17a}. The first of these mosaics targeted the OMC-1 region (blue footprints in Fig.~\ref{fig:ISF_VISTA} right) covering the central region of the ONC, including the Trapezium and the Orion BN source, the Orion BN/KL explosion \citep{BAL11}, the OMC-1 South proto-cluster \citep{GRO05}, the OMC-1 ridge \citep[][also referred to as OMC-1N]{WIS98}, and some of most prominent dense molecular fingers \citep{ROD92}. Continuing to the north, the second mosaic mapped the OMC-2 and the southern end of the OMC-3 regions (red footprints in Fig.~\ref{fig:ISF_VISTA} right) covering all the previously identified FIR sources \citep[OMC-2 FIR 1-6; ][]{MEZ90} and several of the millimeter sources \citep[MMS 8-10; ][]{CHI97} within these two clouds. 

As primary target line of this project, we observed the N$_2$H$^+$ (1-0) line emission in Band 3 \citep[93173.764 MHz,][]{PAG09} at high spectral resolution (30 kHz or 0.1 km~s$^{-1}$). Three additional broad-band, 1.8~GHz wide spectral windows were observed simultaneously covering the continuum centred at $\sim$~93, 104, and 106 GHz. The observations were carried out with PWV=2-5~mm and in the most compact configuration of the ALMA 12m array (C36-1 and C36-2) with baselines between 15.1 and 310 meters. The quasar J0423-0120 was observed for bandpass plus amplitude calibrations at the beginning of each observing block. Phase calibration was performed on J0541-0541 every $\sim$~10min. Independent visibility data for each mosaic were obtained in CASA (v4.5.3) \citep{CASA07} using the facility provided pipeline. 

Our two N$_2$H$^+$ ALMA mosaics were simultaneously imaged in CASA (v4.7.1) in combination with previous single-dish IRAM~30m observations of the ISF \citep[$\theta_{FWHM}=30"$,][]{HAC17a} using standard techniques. First, we subtracted the continuum emission from our high resolution spectral window based on the line-free continuum level estimated in all sidebands. Second, the line visibilities were simultaneously deconvolved with the CASA task {\it clean} using the single-dish observations as source model and a Briggs weighting with robust parameter equal to 0.5. Third, the resulting primary-beam corrected image (at $\sim3.5"\times3.0"$ resolution) was convolved into a final circular beam ($\theta_{FWHM}$) of 4.5" in order to improve the sensitivity and stability of our maps. Finally, both single-dish and interferometric maps were combined using the task {\it feathering} in order maximize the recovery of the extended emission filtered by the 12m array. The rms level of our final map,  estimated  from the analysis of line-free channels, is 25 mJy~beam$^{-1}$ at a spectral resolution of 0.1 km~s$^{-1}$. Assuming a flux conversion factor of $13.6\left(\frac{300GHz}{\nu}\right)^2 \left(\frac{1"}{\Theta_{FWHM}}\right)^2 = 6.96$ K~Jy$^{-1}$ (ALMA technical handbook), the above estimates translate into a brightness temperature sensitivity of 0.17~K in main beam units. 
The recovered signal in our spectra covers a wide dynamic range in intensities with peak values up to S/N~$\gtrsim$~50 with respect to the noise levels in both total integrated emission and individual line intensities.
A detailed discussion on the data reduction process will be presented in a subsequent paper (Paper II; Hacar et al in prep.).


\section{Results}\label{sec:densegas}

We investigated the internal gas substructure of the ISF from the analysis of the N$_2$H$^+$ (1-0) line emission.
Enhanced as result of the CO freeze-out, N$_2$H$^+$ is an ideal tracer of cold gas with densities of n(H$_2$)~$\gtrsim 5\times 10^{4}$ cm$^{-3}$ \citep{BER07}.
The emission of its ground transition J=(1-0) has been traditionally employed in observations of dense cores \citep[e.g.,][]{CAS02}.
Recent studies of massive star-forming regions indicate that the emission of this tracer is not restricted to these stellar embryos but that it extends to large scales in dense environments \citep{FER14,HEN16,HAC17b}. In the case of the ISF, single-dish observations indicate a widespread and intense N$_2$H$^+$ emission along the main axis of this massive filament \citep{TAT08,HAC17a}.

On the other hand, N$_2$H$^+$ also presents several observational advantages for star-formation studies.
Unlike the FIR/mm continuum observations (e.g., {\it Herschel}) sensitive to the total column density, N$_2$H$^+$ selectively highlights the high-density, star-forming material in compact regions \citep[e.g.,][]{PET16,HAC17b}.
Moreover, its emission properties and hyperfine structure enable high accuracy studies of the gas kinematics otherwise hampered by the more complex hyperfine structure in other dense tracers like ammonia \citep[see a discussion in][]{HAC17b}.  

\begin{figure*}
	\centering
	\includegraphics[width=0.93\textwidth]{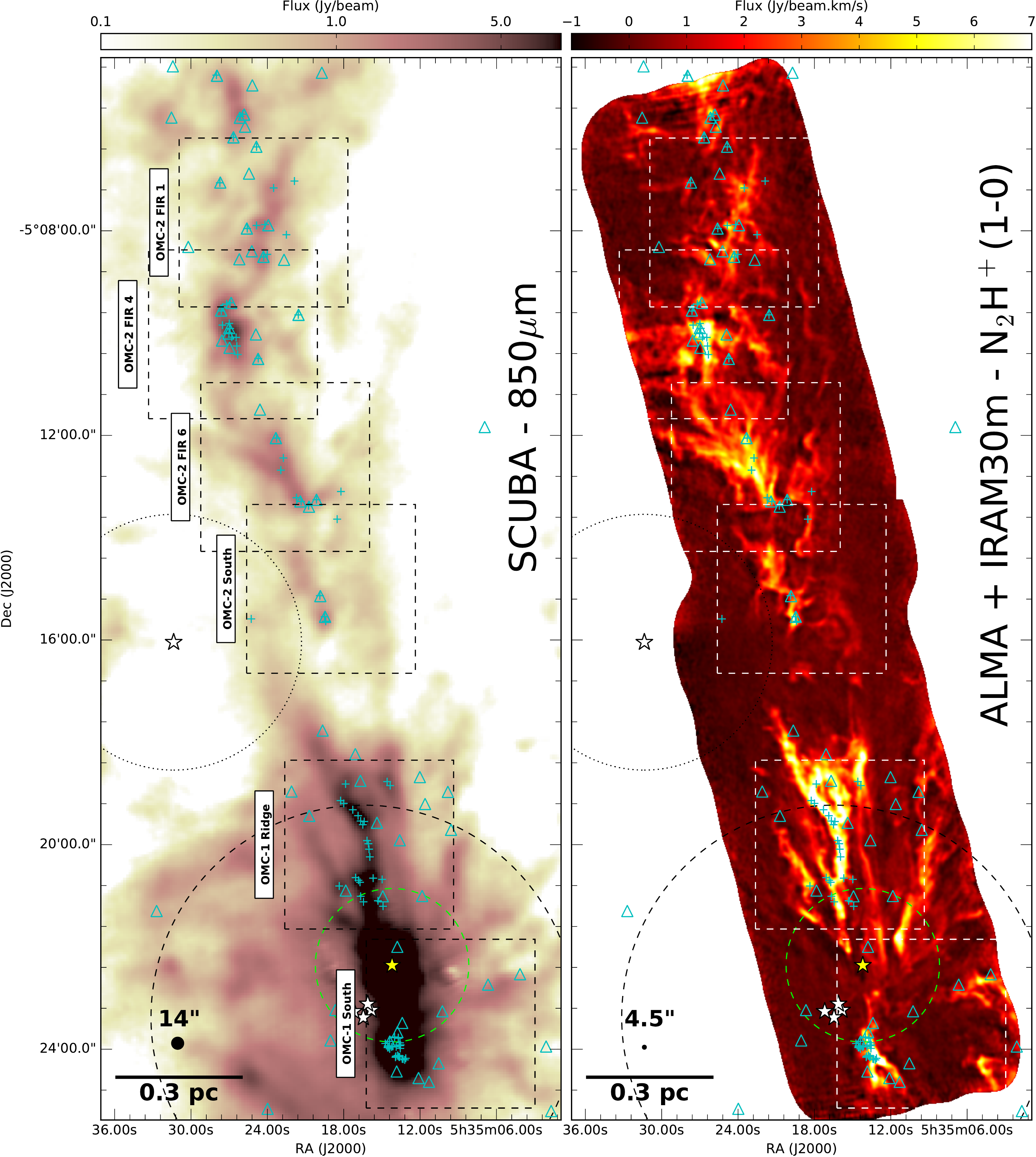}
	\caption{Dense gas distribution within the ISF. 
		{\bf (Left)} SCUBA-850~$\mu$m continuum emission \citep{JOH99}. {\bf (Right)} Total N$_2$H$^+$ integrated intensity mosaic obtained by the combination of ALMA~12m and IRAM~30m data.
		For reference, the positions of the Trapezium (white stars in OMC-1), the Orion BN source (yellow stars), the size of the Orion BN/KL explosion \citep[green dashed circle;][]{BAL11}, the 0.5~pc radius of the ONC (dashed line), the extension of the M43 nebula \citep[dotted circle;][]{SUB01} powered by NU Ori (isolated white star), and both Spitzer protostars \citep[blue triangles;][]{MEG12,STU13,FUR16} plus continuum sources \citep[blue crosses;][]{TEI16,KAI17,PAL17} are shown in both panels. 
		The corresponding beamsize (black solid dot) is indicated in the lower left corner in comparison with a characteristic 0.3~pc scale (black bar). The position of the zoom-in regions presented in Fig.~\ref{fig:ISF_regions} are enclosed by dashed boxes.
		A movie showing the combined ALMA~12m plus IRAM~30m mosaic is included as on-line material.
	}
	\label{fig:ISF_ALMA_SCUBA}
\end{figure*}

Figure~\ref{fig:ISF_ALMA_SCUBA} (right) shows the total integrated intensity N$_2$H$^+$ (1-0) emission map obtained after the combination of our IRAM~30m and ALMA observations. The resulting ALMA mosaic covers an approximate area of 2.5~$\times$~0.48~pc$^2$ with an effective resolution of 0.009~pc \citep[i.e.,  $\theta_{FWHM}=4.5"$ at the distance of 414~pc;][]{MEN07}. Several prominent gas concentrations are coincident with the position of the OMC-1 South proto-cluster, the OMC-1 Ridge region, and the OMC-2 FIR 1, 4, plus 6 sources with flux densities $>$~3~Jy~beam$^{-1}$~km~s$^{-1}$. In addition, a rich filamentary substructure can be identified at lower intensities.
Among them, the well-known molecular fingers in the OMC-1 region \citep{ROD92} are clearly seen in these observations. 

Our new ALMA mosaic allows us to investigate the distribution of dense and star-forming gas in the ISF with unprecedented detail.
As illustrated in Fig.~\ref{fig:ISF_ALMA_SCUBA}, we find a close correspondence between the recovered total N$_2$H$^+$ intensity and the high column density material traced in previous 850~$\mu m$-SCUBA observations \citep[][ $\theta_{FWHM}=$14.0'']{JOH99}.
Similarly, the N$_2$H$^+$ emission features mimic the dense gas distribution reported in interferometric VLA-NH$_3$ maps of OMC-1 \citep[][ $\theta_{FWHM}=8.5" \times 9"$, $\delta v=0.3$~km~s$^{-1}$]{WIS98} and OMC-2 \citep[][ $\theta_{FWHM}=5"$, $\delta v=0.6$~km~s$^{-1}$]{LI13}. With the only exception of the hot gas component at the vicinity of the Orion BN/KL region \citep[T$_K>$~100~K;][]{GEN82,GOD11}, N$_2$H$^+$ accurately traces the dense and cold material (T$_K\lesssim$~30~K) in this cloud.
The N$_2$H$^+$ integrated emission also encloses $>$~90\% of the {\it Spitzer} protostars \citep[blue triangles;][]{MEG12,STU13,FUR16} and compact continuum sources \citep[blue crosses;][]{KAI16,TEI16,PAL17}.
In addition to these previous works, the improved spatial and spectral resolutions of our ALMA observations ($\theta_{FWHM}=4.5"$, $\delta v=0.1$~km~s$^{-1}$) enable an accurate characterization of the dense gas properties both in space and velocity.

\begin{figure*}
	\centering
	\includegraphics[width=0.92\textwidth]{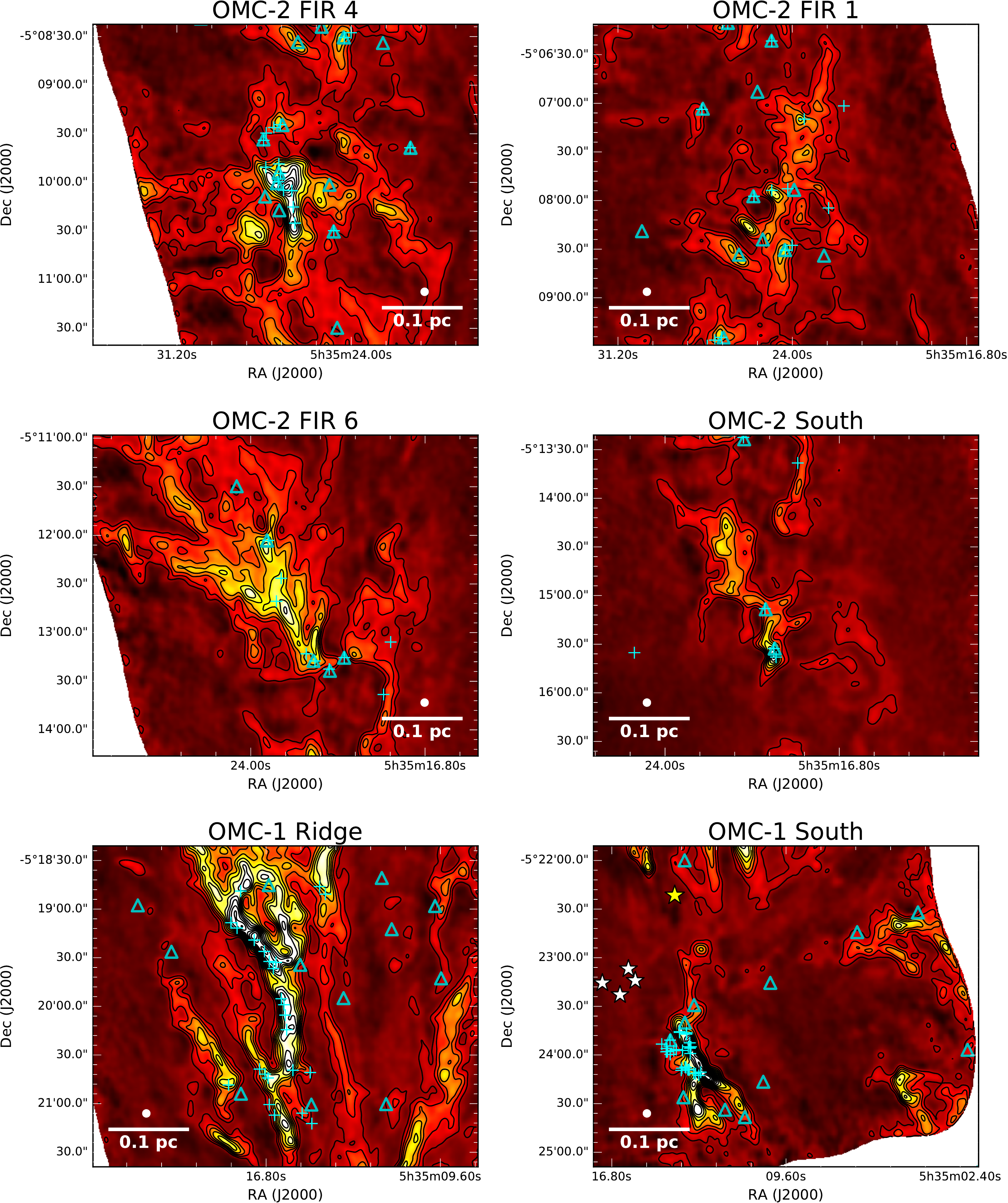}
	\caption{Close-up view of the total N$_2$H$^+$ integrated emission in 6 selected regions along the ISF (see also Fig.~\ref{fig:ISF_ALMA_SCUBA}). from left to right and from top to bottom: (a) OMC-2 FIR~4; (b) OMC-2 FIR~1; (c) OMC-2 FIR~6; (d) OMC-2 South; (e) OMC-1 Ridge; and (f) OMC-1 South. The positions of all the Spitzer protostars \citep[blue triangles;][]{MEG12,STU13,FUR16} and continuum sources \citep[blue crosses;][]{TEI16,KAI17,PAL17} are indicated in all subpanels. Contours are equally spaced every 1~Jy~beam$^{-1}$~km~s$^{-1}$. The beamsize (white solid dot) and the characteristic 0.1~pc scale (white bar) are indicated in the lower corner of each subpanel.
	}
	\label{fig:ISF_regions}
\end{figure*}

The high dynamic range of our N$_2$H$^+$ ALMA+IRAM~30m observations reveals the intricate gas substructure of the massive ISF. 
As highlighted in Fig.~\ref{fig:ISF_regions} (see boxes), the most prominent clumps identified in previous studies split into multiple independent elongated substructures at sub-parsec scales, clearly separated at the resolution of our ALMA observations (e.g., see OMC-2 South or OMC-2 FIR-1). Each of these substructures branches off into smaller filamentary features below 0.1~pc. At smaller scales, these latter objects seem to fragment, forming series of prolate condensations with major axes between $\sim$~0.01-0.03~pc. 
Despite their differences in terms of star-formation activity and feedback, the same gas organization is simultaneously observed in both OMC-1 and OMC-2 regions (see also Sect.~\ref{sec:environment}).
Within the boundaries of our maps, this substructure extends over (at least) two orders of magnitude in scale between 0.02 and $\sim$~2~pc.

The above hierarchical organization of {\it filaments within filaments} resembles the so-called {\it bundles of fibers} identified in low-mass filaments \citep{HAC13,HAC16} and intermediate-mass clusters \citep{FER14,HAC17b}. Our ALMA observations demonstrate that the existence of these fiber networks is not restricted to low-mass regions but extends to filaments at higher mass regimes like the ISF. In the following subsections we investigate the main physical properties of these new ISF fibers in more detail.

\subsection{Fibrous substructure of the ISF}\label{sec:fiberID}
 
We have characterized the internal gas substructure of the ISF using a new version of the FIVE analysis technique \citep{HAC13}, hereafter referred to as HiFIVE. A description of the algorithm and its performance can be found in Appendix~\ref{sec:HiFIVE}. In brief, HiFIVE uses a new hierarchical scheme to systematically identify and characterize velocity-coherent structures in complex molecular line datasets with large dynamic range and highly variable velocity fields.  HiFIVE carries out this analysis from the continuity of the gas velocity centroids both in space and velocity using a linking velocity gradient adapted to the local line properties.
We analysed more than 70~000 spectra in our data to find $\sim$~25~000 components with S/N~$\ge$~3 (Table~\ref{table:dense_prop}; see also Appendix~\ref{sec:fit_spectra}).
Using HiFIVE, we identify a total of 55 velocity-coherent elongated fibers along the ISF. Our results reveal the extraordinary fibrous nature of the dense gas within this massive filament. Still partially unresolved in our HiFIVE analysis, these 55 fibers should be interpreted as a first order description of the real dense gas substructure in the ISF (see Appendix~\ref{sec:HiFIVE} for a discussion).

\begin{figure*}
	\centering
	\includegraphics[width=0.93\textwidth]{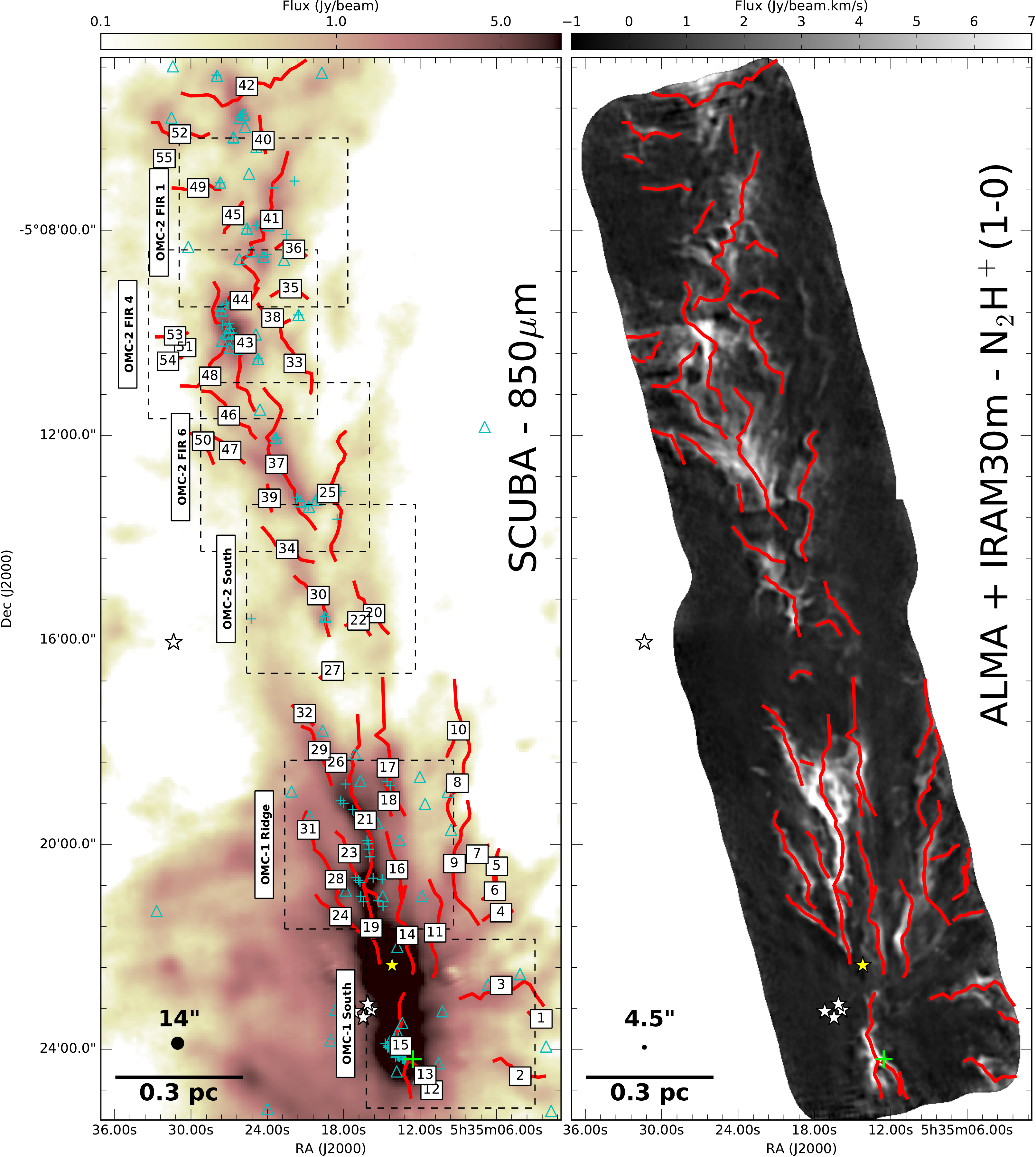}
	\caption{Main axis of the 55 fibers identified by our HiFIVE analysis along the ISF (red segments). 
		{\bf (Left)} SCUBA-850~$\mu$m continuum emission \citep{JOH99}. {\bf (Right)} Total N$_2$H$^+$ integrated emission. The position of both protostars (blue triangles) and continuum sources (blue crosses) are indicated similar to Fig.~\ref{fig:ISF_ALMA_SCUBA}.
		For reference, the positions of the Trapezium and NU Ori stars (white stars) plus the Orion BN source (yellow stars), are highlighted in both panels. 
		The corresponding beamsizes are indicated in the lower left corner in comparison with a characteristic 0.3~pc scale (black bar).  The position of the zoom-in regions presented in Fig.~\ref{fig:ISF_regions} are enclosed by dashed boxes in the SCUBA map.
		We notice the good correspondence between the position of fibers and the location of the intensity enhancements in the continuum. 
		A green cross indicates the positions with the most blue-shifted velocity detected in our N$_2$H$^+$ maps (see also Appendix~\ref{sec:fit_spectra:kinematics}), adopted as the approximate centre of collapse of the OMC-1 region \citep{HAC17a}.
	}
	\label{fig:ISF_ALMA}
\end{figure*}

We have defined the main axis of each individual fiber (red segments in Fig.~\ref{fig:ISF_ALMA}) using the same fitting procedure introduced in \citet{HAC13}. Reinforcing the similarities with previous studies, many of the ISF fibers appear to be well isolated in space and can be recognized in the integrated intensity maps showing large aspect ratios (e.g., fibers \# 25 or 41). 
In Table~\ref{table:fiber_prop}, we summarize the average fiber properties calculated along their main axes. Also displayed in Fig.~\ref{fig:fiber_histo} (left), the ISF fibers present a well-constrained distribution of sizes with an average total length of $0.16\pm0.10$~pc (orange filled histogram) without correcting for projection effects. 

The observed fiber substructure accurately reproduces the internal dense gas distribution of the ISF.
Among the total 295~M$_{\sun}$ of dense gas detected in our high-S/N N$_2$H$^+$ spectra, 288~M$_{\sun}$ (i.e. 98\%) are recovered as fibers (see the conversion between the total N$_2$H$^+$ integrated intensities and total column densities in Sect.~\ref{sec:stability}). 
In most cases, we notice a correspondence between these N$_2$H$^+$ fibers and the intensity enhancements detected in the continuum (see Fig.~\ref{fig:ISF_ALMA}, right panel).
Indeed, the vast majority of compact sources and protostars in the ISF are found in association to these fibers (e.g., fibers \# 21, 37, 43, etc). This complex fiber distribution entirely determines the internal organization of the ISF.

Overall, we find no correlation between the orientation of fibers and large-scale feedback effects in the ISF (see also Sect.~\ref{sec:environment}).
The fibers are distributed irrespective of the stellar activity within the cloud (e.g., see ONC vs. OMC-2 FIR 1).  
While originally independent, some of the fibers might still be locally influenced by the presence of stars.
Within the OMC-1 region, fibers are radially oriented pointing towards the OMC-1 South proto-cluster, 
likely reflecting the global gravitational collapse of this cloud \citep[][see also Appendix~\ref{sec:fit_spectra:kinematics}]{HAC17a}.
The extension of these fibers beyond the Orion BN/KL explosion \citep[green dashed circle in Fig.~\ref{fig:ISF_ALMA_SCUBA}][]{BAL11} rules out a direct connection with this energetic event. 
In some specific cases, however, the individual fiber morphology might be potentially altered by the local influence of stars (e.g., fibers \#25 \& \#37), and both the M43 nebula (e.g., fibers \#30 \& \#34) or the ONC (e.g., fibers \#19 \& 24) (see also Fig.~\ref{fig:ISF_ALMA_SCUBA}).
In spite of these localized effects,
the reported fiber organization appears to reflect the original gas substructure before the formation of stars.

\subsection{Kinematic properties: subsonic fibers in massive clouds}\label{sec:sigmaNT}

Investigating the magnitude of the line-of-sight (l.o.s.) non-thermal velocity dispersion $\sigma_{NT}$ is of fundamental importance to characterize the internal dynamical state of fibers \citep[e.g.,][]{HAC11}. 
This observable can be estimated from the measured line full-width-half-maximum ($\Delta V$) obtained from our hyperfine fits (see Appendix~\ref{sec:HiFIVE}):
\begin{equation}\label{eq:sigmaNT}
	\sigma_{NT}=\left[\left(\frac{\Delta V}{\sqrt{8 ln~2}}\right)^2-\frac{k_B T_\mathrm{K}}{\mu(N_2H^+)} \right]^{1/2}.
\end{equation}
Compared to the (local) thermal sound speed for H$_2$,  $\mathrm{c}_s(T_\mathrm{K})=\sqrt{\frac{k_B T_\mathrm{K}}{\mu(H_2)}}$, the ratio $\sigma_{NT}/\mathrm{c}_s(T_\mathrm{K})$ can be used as a diagnostic tool to determine whether the observed gas motions are subsonic ($\sigma_{NT}/\mathrm{c}_s(T_\mathrm{K})\le$~1), transonic (1~$ < \sigma_{NT}/\mathrm{c}_s(T_\mathrm{K})\le$~2), or supersonic ($\sigma_{NT}/\mathrm{c}_s(T_\mathrm{K})>$~2).

In the absence of strong feedback effects, the dense gas in clouds is found at low and relatively uniform temperatures, typically at T$_K\sim$~10~K \citep{MYE83}. Based on this property, the dynamical state of fibers in quiescent environments is commonly evaluated assuming a single H$_2$ sound speed $\mathrm{c}_s(T_\mathrm{K})\sim\mathrm{c}_s(10\ K)$ \citep[e.g.,][]{HAC13,HAC17b}.
Contrary to low mass clouds, the NH$_3$-derived T$_K$ values in the vicinity of the ONC indicate large thermal variations leading to $\mathrm{c}_s(T_\mathrm{K})\gg\mathrm{c}_s(10\ K)$  \citep[e.g., see][]{WIS98}. To correctly evaluate these thermal effects, we combined our new ALMA observations with the NH$_3$ T$_K$ estimates provided by the GAS-DR1 survey \citep[$\theta_{FWHM}=32"$; see details in][]{FRI17}. 
As a first order approximation, we adopt
the T$_K$ value for each individual N$_2$H$^+$ component from the nearest position surveyed in NH$_3$ \footnote{We note that the GAS-NH$_3$ survey provides a unique temperature per position. When multiple N$_2$H$^+$ components are identified in a single ALMA spectrum, the same T$_K$ value is assigned to all of them.
}. 
With similar density regimes traced by these twin molecules \citep[see][]{HAC17b}, these NH$_3$-derived temperatures provide good estimates for the thermal state of the dense gas detected in N$_2$H$^+$. 
In Table~\ref{table:dense_prop} we list the average T$_K$ estimates and their corresponding sound speed $\mathrm{c}_s(T_\mathrm{K})$ values in our maps.

\begin{figure*}
	\centering
	\includegraphics[width=\linewidth]{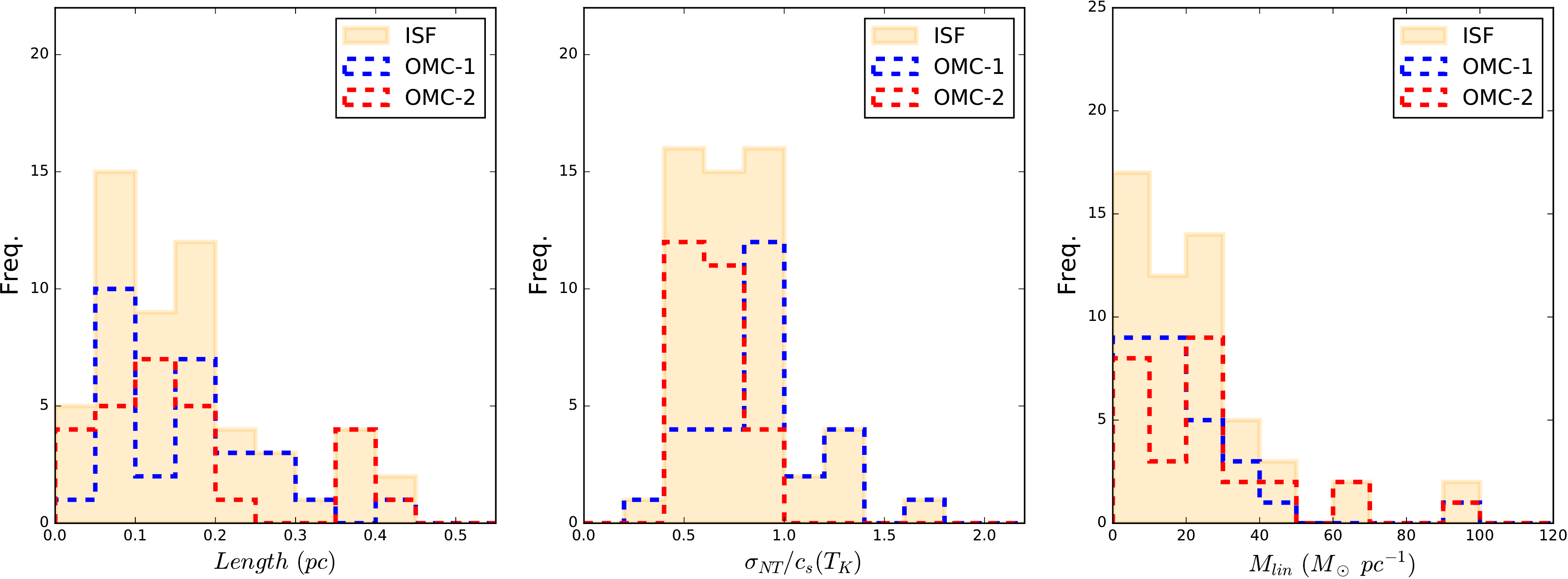}
	\caption{Statistical properties of the ISF fibers (orange shaded area): {\bf (Left)} Total fiber length; {\bf (Centre)} average line-of-sight, non-thermal velocity dispersion $\sigma_{NT}$ as a function of the local sound-speed c$_s(T_\mathrm{K})$; and {\bf (Right)} total mass-per-unit length.
		The individual fiber properties in OMC-1 and OMC-2 are highlighted by blue and red dashed lines, respectively.
	}
	\label{fig:fiber_histo}
\end{figure*}

%
\begin{table}
	\caption{Dense gas properties}             
	\label{table:dense_prop}      
	\centering                          
	\begin{tabular}{c c c c}        
		\hline\hline                 
		& ISF & OMC-1 & OMC-2 \\    
		\hline
		N$_2$H$^+$ fits (S/N~$\ge$~3) 									& 25078		& 13305    & 11773 \\
		T$_K$ (K)$^{(a)}$			& 24.8$\pm$10.3 	& 29.0$\pm$12.5    	& 20.1$\pm$2.2 \\
		
		$\left<\mathrm{c}_s (\mathrm{T}_K)\right>$~(km~s$^{-1}$)			& 0.294 	& 0.318    	& 0.267 \\
		
		$\Delta V$~(km~s$^{-1}$)								& 0.607		& 0.722    & 0.477 \\
		$\left<\sigma_{NT}\right>$~(km~s$^{-1}$)			& 0.240  	& 0.288    & 0.185 \\
		$\left<\sigma_{NT}/\mathrm{c}_s(T_\mathrm{K})\right>$			& 0.81 	& 0.91    & 0.69 \\
		\hline
		$\sigma_{NT}/\mathrm{c}_s(T_\mathrm{K})\le$~1					& 76.4\%	& 68.4\%   & 85.5\% \\
		1~$ < \sigma_{NT}/\mathrm{c}_s(T_\mathrm{K})\le$~2			& 20.0\% & 25.7\%	& 13.6\% \\
		$\sigma_{NT}/\mathrm{c}_s(T_\mathrm{K})>$~2 					& 3.6\%		& 5.9\%   & 0.9\% \\
		\hline                                  
	\end{tabular}
	\tablefoot{
		\tablefoottext{$a$}{Ammonia-derived gas kinetic temperatures obtained from the GAS survey \citep{FRI17}.}
	}
\end{table}

We summarize the 
statistical properties of the derived $\sigma_{NT}$ measurements obtained using Eq.~\ref{eq:sigmaNT} for the all the N$_2$H$^+$ components detected with S/N~$\ge$~3 in Table~\ref{table:dense_prop}.
This table also includes the fraction of components in different turbulent regimes in the ISF. Overall, the internal gas motions within this filament are described by an average non-thermal velocity dispersion of $\left<\sigma_{NT}/\mathrm{c}_s(T_\mathrm{K})\right>=0.81$. Remarkably, $>$~75\% of the positions detected in N$_2$H$^+$ show subsonic velocity dispersions (i.e., $\sigma_{NT}/\mathrm{c}_s(T_\mathrm{K})\le$~1). Conversely, less than 4\% of the N$_2$H$^+$ components exhibit supersonic motions  (i.e., $\sigma_{NT}/\mathrm{c}_s(T_\mathrm{K})>$~2). Our analysis therefore indicates that most of the dense gas within the massive ISF is relatively quiescent, typically showing subsonic ($>$~75\%) or subsonic+transonic ($>$~95\%) non-thermal motions.
 
\begin{figure}
	\centering
	\includegraphics[width=\linewidth]{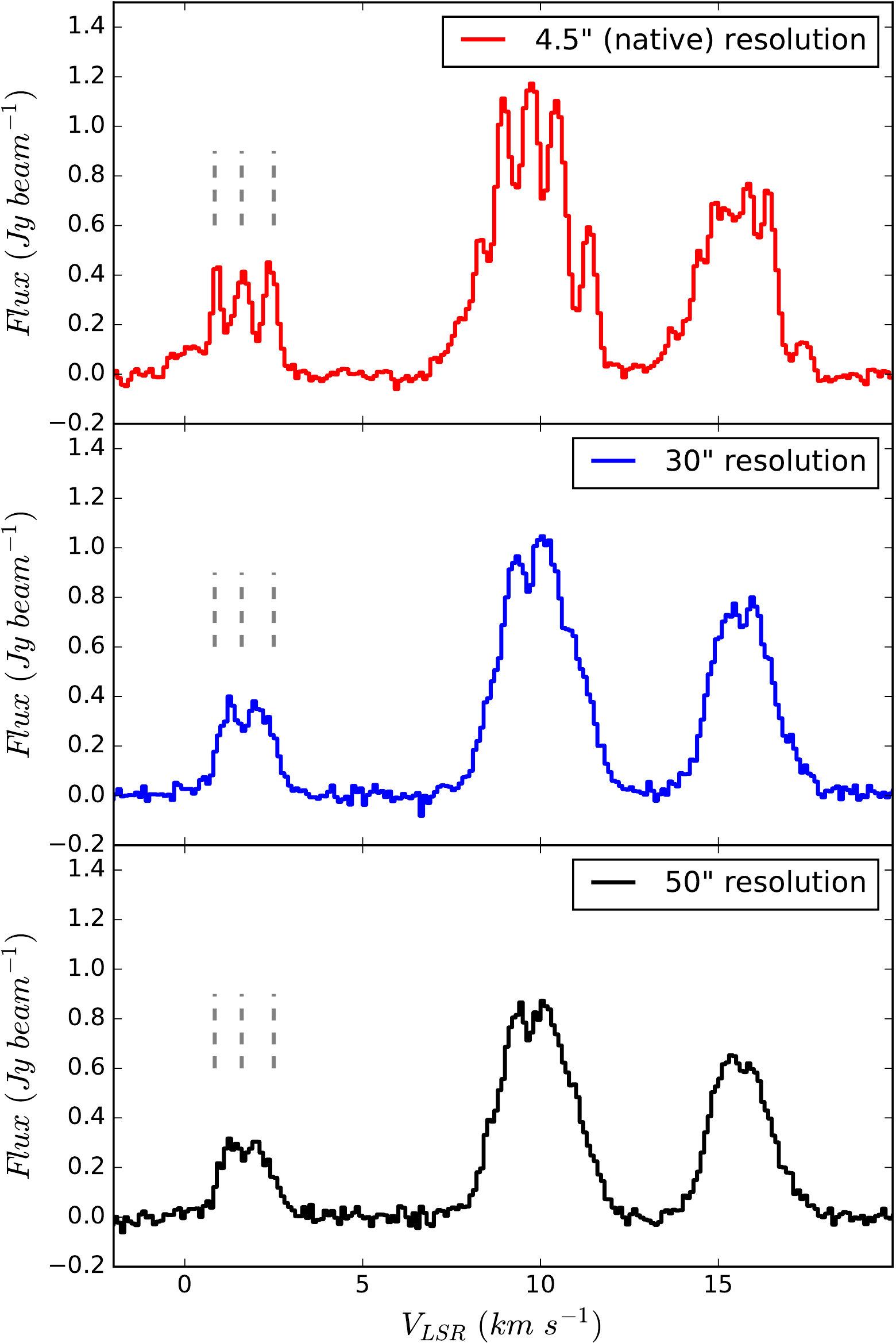}
	\caption{Resolution effects affecting the identification of spectral line components at
		different resolutions: $\theta_{FWHM}=$~4.5" (native; upper panel), $\theta_{FWHM}=$~30"  (mid panel), and $\theta_{FWHM}=$~50" (lower panel). For comparison, all spectra include the same noise level.
		We note how the three original velocity components, clearly detected in the N$_2$H$^+$ isolated hyperfine line at the native resolution (grey dashed lines), are progressively diluted and blended when convolved into larger beams. In particular, we note that most of the original line multiplicity is lost at the typical resolution of single-dish observations \citep[mid panel; e.g.][]{HAC17b}.
	}
	\label{fig:spectra}
\end{figure}

In Fig.~\ref{fig:fiber_histo} (centre), we show the histogram for the mean $\sigma_{NT}$ values (in units of the corresponding local $\mathrm{c}_s(T_\mathrm{K})$; orange filled histogram) for all the fibers extracted by HiFIVE. The ISF fibers exhibit non-thermal velocity dispersions between 0.4 and 1.5 times their local sound speed.
Low, sonic-like internal motions have been reported as an intrinsic characteristic of fibers in previous studies \citep{HAC11,HAC13,HAC17b}. This unique property is also shared by the ISF fibers, where all the fibers detected in this massive filament are dominated by (tran-)sonic internal motions.

The small velocity dispersions measured in the ISF fibers appear to be in contradiction with the large linewidths reported in active regions like OMC-1 \citep[e.g., ][]{FRI17}.  Our observations suggest that most of these broad emission lines detected in tracers like N$_2$H$^+$ \citep{TAT08} and NH$_3$ \citep{FRI17} are the result of a combination of multiple gas components and local velocity gradients typically unresolved within a single-dish beam. We illustrate this effect in Fig.~\ref{fig:spectra} by comparing the emission in a single position recovered using different effective beam sizes. At the native ALMA resolution (red spectrum; upper panel), the observed N$_2$H$^+$ emission shows three independent narrow lines, clearly separated in the hyperfine isolated component (grey dashed segments). Due to the complex gas kinematics within this region, the above line substructure is progressively smeared out when convolved with neighbouring positions to the resolution of previous single-dish studies (e.g., blue spectrum; central panel). Dilution and blending effects give the appearance of a single broad component at even larger beam sizes (black spectrum; lower panel). The above comparisons highlight the importance of both spectral and spatial resolution in the kinematic analysis of massive clouds. 

\subsection{Linear masses and stability}\label{sec:stability}

\begin{figure}
	\centering
	\includegraphics[width=\linewidth]{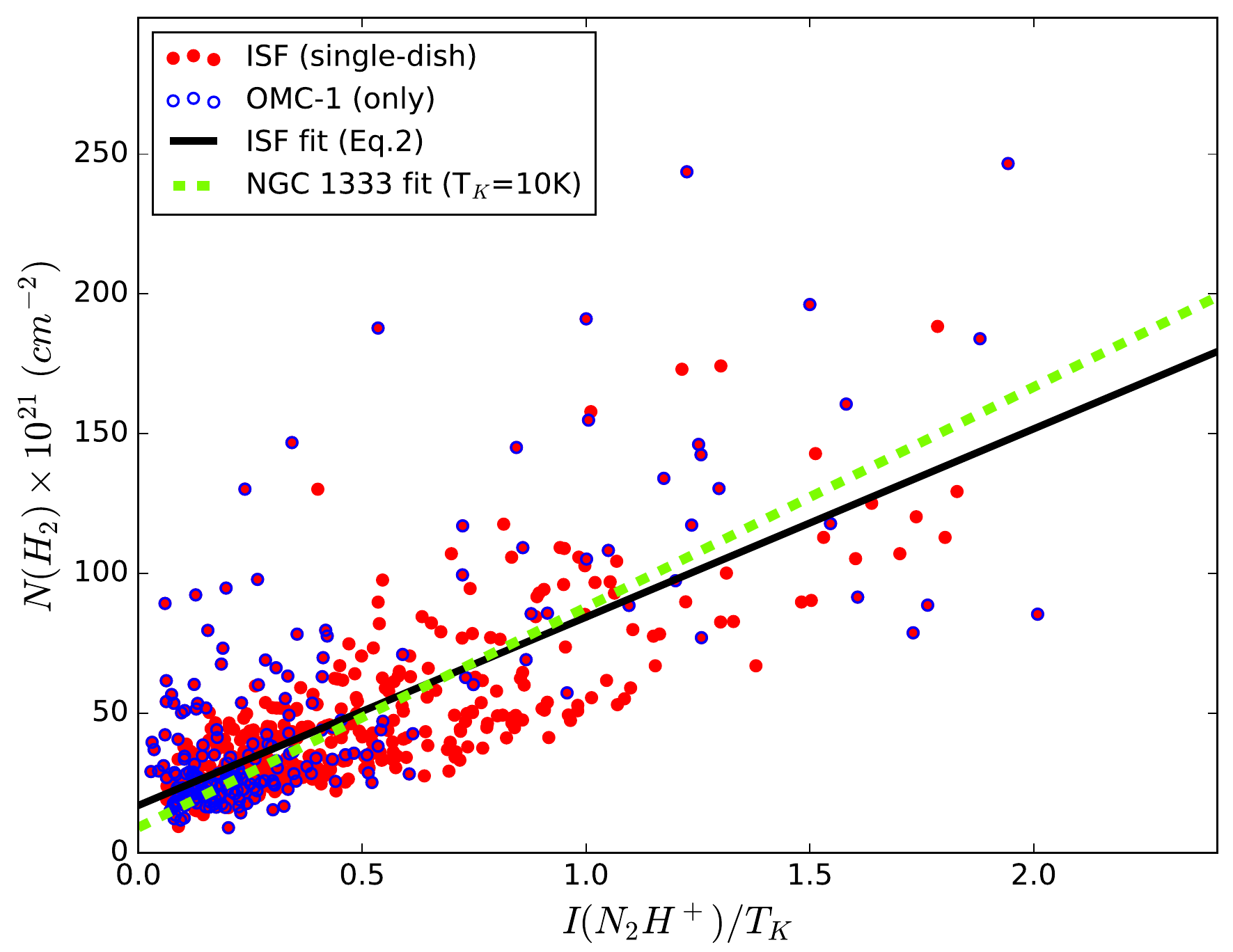}
	\caption{Empirical correlation between the observed single-dish N$_2$H$^+$ intensities \citep{HAC17a}, normalized by the local gas kinetic temperature T$_K$ \citep{FRI17}, and the total (gas+dust) column densities N(H$_2$) derived in previous {\it Herschel-Planck} measurements \citep{LOM14}. This plot includes all the positions detected at $R$(Trapezium)~$\ge$~0.3~pc in both OMC-1 and OMC-2 clouds (red dots). Those points belonging to the OMC-1 region are highlighted in blue. The thick black line indicates the results of the linear fit defining Eq.~\ref{eq:colden}. The green dashed line indicates the expected correlation for the observed N$_2$H$^+$ intensities in the NGC1333 proto-cluster assuming a constant temperature of 10~K \citep{HAC17b}.
	}
	\label{fig:N2H+_cal}
\end{figure}

Characterizing the stability of the ISF fibers requires the study of their internal mass distribution. In the absence of additional line information, we have calibrated our N$_2$H$^+$ emission with previous {\it Herschel} surveys along the ISF region. Introduced in similar studies using N$_2$H$^+$ as line tracer \citep{TAF15,HAC17b}, this method obtains an empirical conversion factor between the observed integrated N$_2$H$^+$ intensities and the equivalent gas plus dust column density. Detailed comparisons with radiative transfer Monte-Carlo simulations prove the validity of this technique in the case of optically thin emission \citep[see also][]{TAF15}.

Figure~\ref{fig:N2H+_cal} illustrates a point-to-point comparison of the total N(H$_2$) column density derived in previous {\it Herschel-Planck} studies \citep[$\theta_{FWHM}=36"$;][]{LOM14} with similar single-dish N$_2$H$^+$ integrated intensity maps \citep[$\theta_{FWHM}=30"$;][]{HAC17b}, normalized by the corresponding gas kinetic temperature from the GAS-NH$_3$ survey \citep[$\theta_{FWHM}=32"$;][]{FRI17} (red filled circles)\footnote{We have excluded those positions within R(Trapezium)~$<$~0.3~pc in Fig.~\ref{fig:N2H+_cal} due to saturation effects in the  {\it Herschel-Planck} maps \citep[see ][]{LOM14}.}. 
With the exception of several noisy positions in OMC-1 (see blue circles), the normalized N$_2$H$^+$ intensities exhibit a roughly linear correlation with the observed {\it Herschel-Planck} total column densities along the ISF. We have described this relationship using a least-squares fit of all the points included in Fig.~\ref{fig:N2H+_cal} (black line) resulting in a linear term:
\begin{equation}\label{eq:colden}
N(H_2)\ [\mathrm{cm}^{-2}]= 67.4\times10^{21}\cdot\left(\frac{I(N_2H^+)\ [\mathrm{K}\ \mathrm{km}\ \mathrm{s}^{-1}]}{T_\mathrm{K}\ [\mathrm{K}]}\right) .
\end{equation}
Within a factor of two, this fit reproduces the observed correlation for column densities between N(H$_2$)~$\sim (20-200)\times10^{21}$~cm$^{-2}$. 
Interestingly, we find an excellent correspondence between Eq.~\ref{eq:colden} in Orion and the results obtained from the study of the N$_2$H$^+$ emission in the NGC~1333 region in Perseus \citep{HAC17b} assuming a uniform T$_K$ of 10~K (i.e., $N(H_2)= 78.8\times10^{21}\cdot\left(\frac{I(N_2H^+)}{10\ K}\right)$; green dashed line). The good agreement between these two independent studies suggest similar abundances for N$_2$H$^+$ in both Orion and Perseus clouds and reinforces the use of this molecule as a robust tracer for dense gas under different physical conditions \citep[e.g.,][]{FOR14,PET16,KAU17}.

We derived the total gas plus dust column density for each of the gas components detected in our ALMA maps from their corresponding integrated N$_2$H$^+$ emission using the empirical correlation described in Eq.~\ref{eq:colden} (see also Appendix~\ref{sec:abundance}). 
The total mass per fiber is then estimated from the addition of all components associated to each individual structure at the resolution of our maps assuming a distance of D=414~pc. Finally, for each of these fibers we derive the mass-per-unit-length M$_{\mathrm{lin}}$ dividing the above total mass by their corresponding length obtained in Sect.~\ref{sec:fiberID}. 
Our length measurements do not consider projection effects making our M$_{\mathrm{lin}}$ values upper limits.
In addition, 
uncertainties of a factor $\sim$~2 are expected for all the above estimates according to the dispersion observed in  Fig.~\ref{fig:N2H+_cal}. The wide variations of the gas and dust properties seen in the vicinity of the ONC lead to larger uncertainties within the OMC-1 cloud.

Several caveats should be considered on the interpretation of our mass-per-unit-length values in Orion.
First, we empirically obtained a unique intensity-to-mass conversion factor (Eq.~\ref{eq:colden})  from the comparison of single-dish and {\it Herschel} surveys.
Extrapolated to the ALMA resolution, our mass conversion assumes that the same correlation applies at different scales and in different gas parcels of this cloud.
Also, and by construction, its calibration is linked to the absolute values and error estimates of the {\it Herschel} dust column densities as well as the gas kinetic temperatures derived at single-dish resolutions.
On the other hand, our mass calculations adopt an optically thin approximation for N$_2$H$^+$ emission in the ISF. Although justified by our opacity estimates in most cases, this assumption can selectively affect the individual masses of several fibers in our sample (see Appendices \ref{sec:fit_spectra:opacity} and \ref{sec:abundance} for a discussion).
Some of the above assumptions are partially responsible of the observed spread in Fig.~\ref{fig:N2H+_cal} and are assumed to be included in the factor 2-3 uncertainties estimated for the slope of Eq.~\ref{eq:colden}. Despite our efforts, larger uncertainties cannot be ruled out, particularly in the OMC-1 fibers. Additional ALMA observations, including multiple transitions and tracers, are needed to better constrain these mass estimates.

As shown in Fig.~\ref{fig:fiber_histo} (right), the derived M$_{\mathrm{lin}}$ values for the ISF fibers typically range between 10 and 60~M$_\sun$~pc$^{-1}$ with a median value of $\sim~19$~M$_\sun$~pc$^{-1}$. Three exceptions are found showing $\mathrm{M}_{lin}>60$~M$_\sun$~pc$^{-1}$, namely, 
fibers \#~21 (OMC-1 ridge; $\mathrm{M}_{lin}=95$~M$_\sun$~pc$^{-1}$),
\#~37 (OMC-2 FIR-6; $\mathrm{M}_{lin}=96$~M$_\sun$~pc$^{-1}$),
 and \#~42 (OMC-2 FIR-4; $\mathrm{M}_{lin}=68$~M$_\sun$~pc$^{-1}$). As discussed in Appendix~\ref{sec:HiFIVE}, these three regions present clear signs of a complex substructure not recovered by our HiFIVE algorithm. Their unusually large mass per-unit-length should be then taken as upper limits of their actual mass properties.

\begin{figure}
	\centering
	\includegraphics[width=\linewidth]{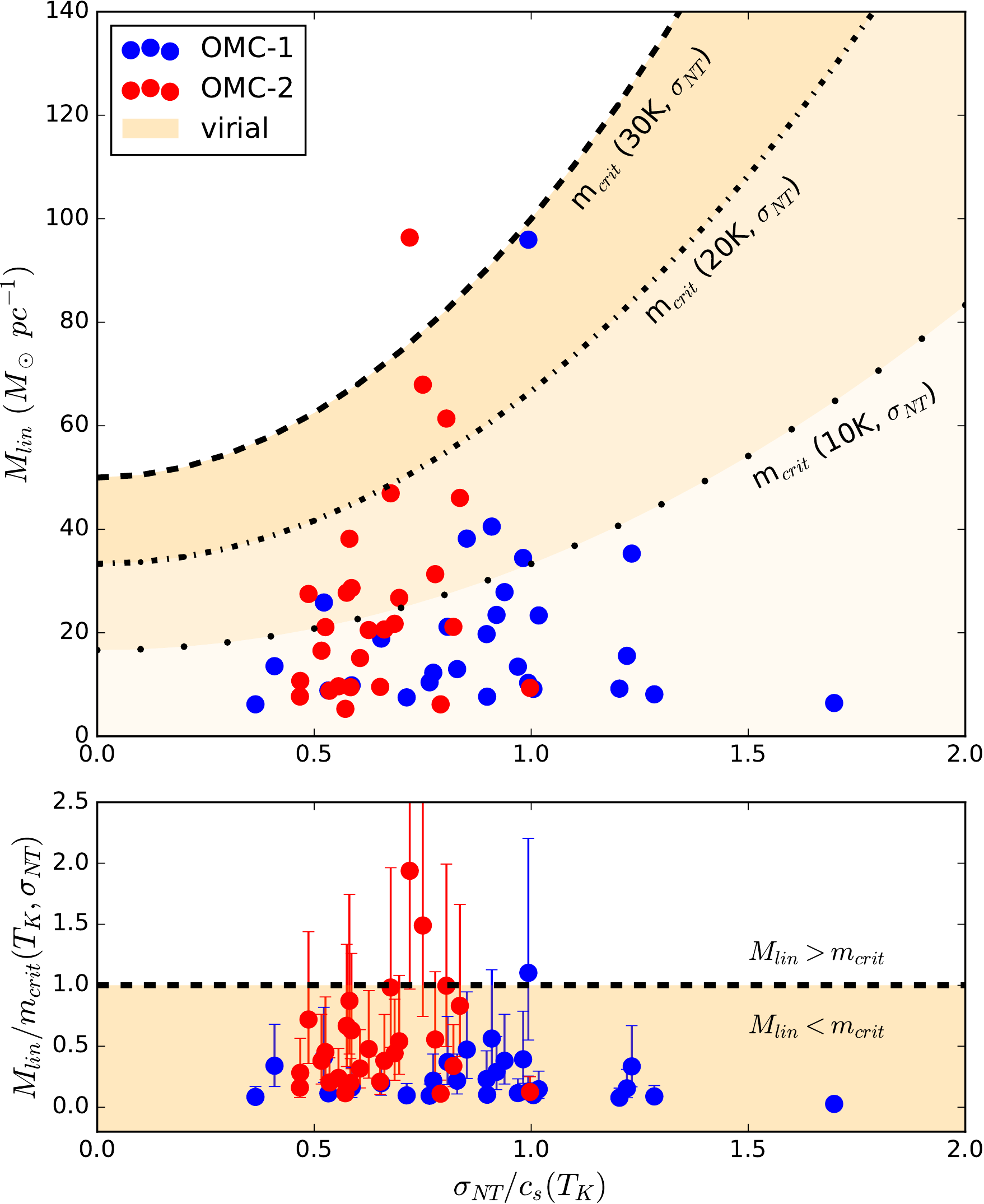}
	\caption{{\bf (Upper)} Mass-per-unit-length (M$_{\mathrm{lin}}$) of the OMC-1 (blue dots) and OMC-2 (red dots) fibers as function of their internal non-thermal velocity dispersion normalized by the local sound speed ($\sigma_{NT}/\mathrm{c}_s(T_\mathrm{K})$). This plot includes the expected critical masses (m$_{crit}$) for an infinite filament in hydrostatic equilibrium at temperatures of 10~K (thick dashed line), 20~K (dashed-dotted line), and 30~K (thick dotted line), respectively (see Eq.~\ref{eq:virial}).
    {\bf (Lower)} Individual critical mass ratio (M$_{\mathrm{lin}}$/m$_{crit}$) for all the fibers in OMC-1 (blue) and OMC-2 (red). For every single fiber the critical mass ratio is calculated relative to its local m$_{crit}(T_K,\sigma_{NT})$. The error bars indicate the factor of two uncertainties associated to our $M_{lin}$ estimates.}
	\label{fig:stability}
\end{figure}

In Figure~\ref{fig:stability} (upper panel), we represent  the dynamical state of the ISF fibers in comparison to their expected masses in equilibrium.
For each individual fiber, we display the observed M$_{\mathrm{lin}}$ as a function of the normalized non-thermal velocity dispersion  $\sigma_{NT}/\mathrm{c}_s(T_\mathrm{K})$ (see Sect.~\ref{sec:sigmaNT}).
We also represent the expected critical mass ($m_{crit}$) for an infinite filament in hydrostatic equilibrium \citep{STO63,OST64}:
\begin{equation}\label{eq:virial}
 m_{crit}(T_\mathrm{K},\sigma_{NT})=\frac{2\ \sigma_{eff}^2}{G}=\frac{2\ \mathrm{c}_s(T_\mathrm{K})^2}{G}\left(1+\left(\frac{\sigma_{NT}}{\mathrm{c}_s(T_\mathrm{K})}\right)^2\right),
\end{equation}
where $\sigma_{eff}^2=\mathrm{c}_s(T_\mathrm{K})^2+\sigma_{NT}^2$ is defined as the effective (thermal + non-thermal) velocity dispersion. 
Typically, $m_{crit}$ is assumed to correspond with the expected value for a thermally supported filament at T$_K=10$~K,  $m_{crit}(\sigma_{eff}=\mathrm{c}_s(10\ K))=16.6$~M$_\sun$~pc$^{-1}$. 
However, this simplified analysis neglects the additional support provided by non-thermal motions and systematically higher temperatures found in active star-forming regions like Orion.
To properly evaluate these effects, in Fig.~\ref{fig:stability} (upper panel) we superpose the evolution of the expected  critical masses for filaments at temperatures of 10~K (dotted line), 20~K (dot-dashed line), and 30 K (dashed line) including the additional $\sigma_{NT}$ contributions in Eq.~\ref{eq:virial}. 
As denoted in this plot, the vast majority of the ISF fibers have subcritical masses (i.e., M$_{lin}\lesssim m_{crit}(T_\mathrm{K},\sigma_{NT})$; shaded areas) for temperatures between 20 and 30~K, similar to those reported in OMC-1 and OMC-2 (see Table~\ref{table:dense_prop}).

To facilitate their comparison, in Figure~\ref{fig:stability} (lower panel) we display the individual ratios of the observed M$_{lin}$ values respect to their local critical mass $m_{crit}(T_K,\sigma_{NT})$ in all the ISF fibers. Each measurement includes a factor of two variation on the M$_{lin}$ estimates (error bars; see above).
In $\sim$~50\% of the cases, the  ISF fibers show (within the errors) critical mass ratios between 0.5~$\le$~M$_{lin}/m_{crit}\le$~1.5. Our results suggest that these fibers are gravitationally bound and largely supported by a combination of thermal and (sonic) turbulent motions presenting a configuration consistent with (or close to) equilibrium. This quasi-stable radial configuration in fibers would provide the necessary conditions for their subsequent fragmentation into cores \citep{INU97,HEI16}.
Possible exceptions to this average behaviour are found in the most massive fibers \#~21, 37, and 42, showing masses potentially exceeding M$_{lin}/m_{crit}\gtrsim$~1.5. While still unresolved in our observations, these high mass ratios suggest that they are gravitationally unstable. 
In contrast, up to $\sim$~50\% of fibers show M$_{lin}/m_{crit}\le$~0.5 and are, therefore, gravitationally unbound (e.g., see some of the OMC-1 fibers).  
In the absence of additional confinement (i.e. thermal pressure, accretion, and/or tidal forces) these latter fibers may be transient objects.

\subsection{Fiber widths: typical 0.03~pc values}\label{sec:fiber_widths}

\begin{figure*}
	\centering
	\includegraphics[width=\textwidth]{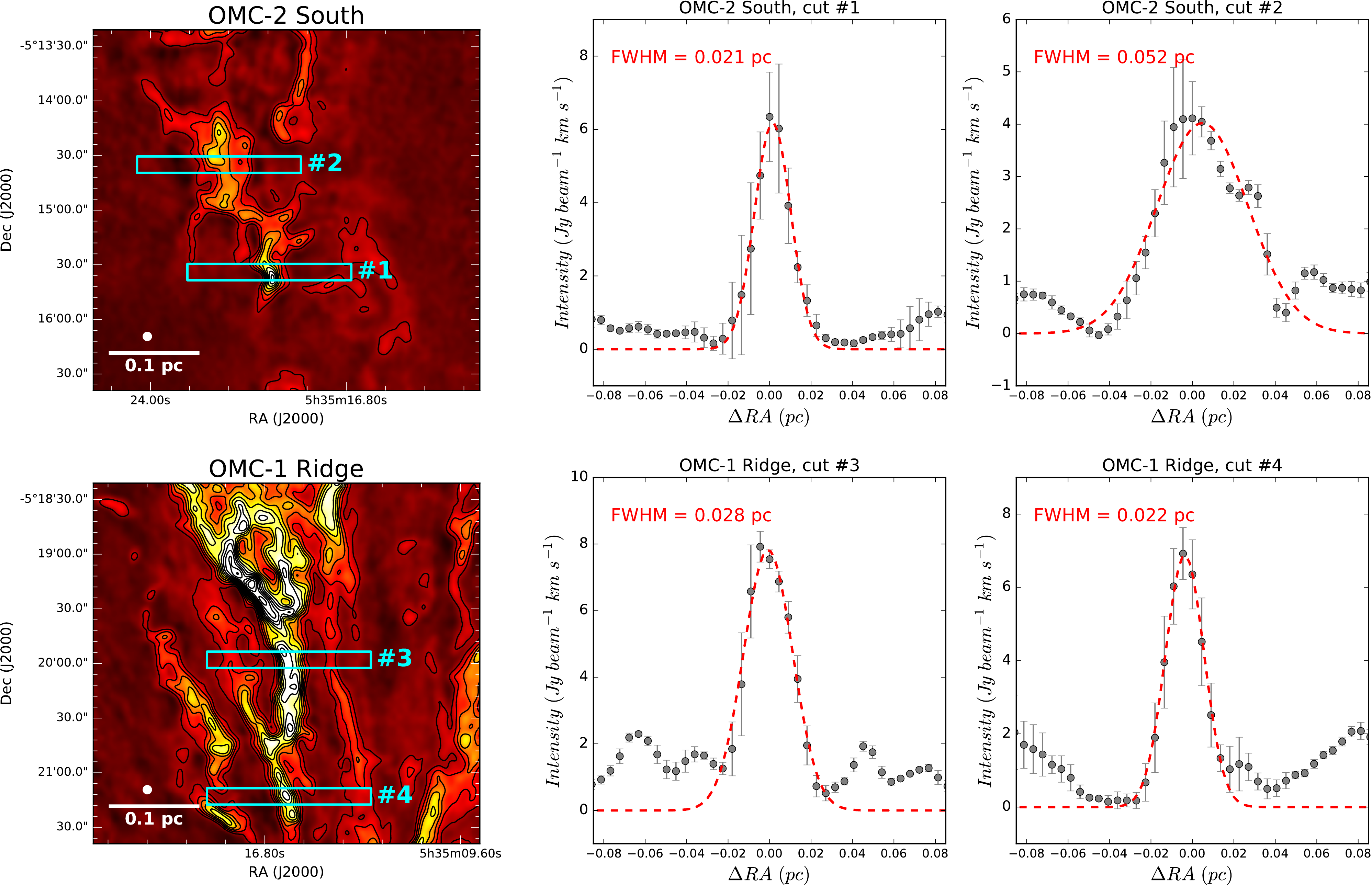}
	\caption{Typical fiber widths in both OMC-2 South (upper panel; fibers \#30 and 34) and OMC-1 Ridge (lower panel; fibers \#19 \& \#21) regions. {\bf (Left subpanels)} Total integrated N$_2$H$^+$ emission including the selected cuts (\#~1-4) perpendicular to the fibers (cyan boxes). Contours are equally spaced every 1~Jy~beam$^{-1}$~km~s$^{-1}$. {\bf (Centre and Right subpanels)} average N$_2$H$^+$ emission along cuts \#~1-4 (grey dots with errors). The red dashed line indicates the Gaussian fits for each profile (see FWHM in the top-left corner of each subplot).
	}
	\label{fig:ISF_cuts}
\end{figure*}

The high-dynamic range of our N$_2$H$^+$ ALMA mosaics reveals a unique characteristic of the ISF fibers. 
As clearly seen in the zoom-in maps of Fig.~\ref{fig:ISF_regions}, most fibers exhibit a compact radial profile showing a sharp emission contrast compared to their local background. 
Several examples of this behaviour can be found in the OMC-1 Ridge, the OMC-1 South, and the OMC-2 FIR-6 regions. Remarkably, all fibers show extremely narrow widths at scales $\ll$~0.1~pc with a sharp emission drop of equivalent column densities $>50$~A$_V$ on scales comparable to our beamsize (i.e., 0.009~pc; see some of the molecular fingers along the OMC-1 Ridge). 
We remind the reader here that the combination of ALMA plus (zero-spacing) single-dish data minimizes potential biases produced by spatial filtering effects (Sect.~\ref{sec:observations}). 
In combination with the optically thin emission properties of N$_2$H$^+$ (Appendix~\ref{sec:fit_spectra:opacity}), the above results describe the intrinsically compact mass distribution within fibers.

In Figure~\ref{fig:ISF_cuts}, we illustrate the ISF fiber widths from the detailed study of several of these objects in both OMC-2 South (upper panel) and OMC-1 Ridge (lower panel) subregions. These fibers were selected for their favourable geometry and orientation, with a clean and well-defined axis.
In each case, we have extracted two horizontal cuts roughly perpendicular to the main fiber axis (cyan boxes in left panels).  
For the selected cuts 1-4, in Figure~\ref{fig:ISF_cuts} (lateral subpanels) we show the corresponding total N$_2$H$^+$ integrated emission as a function of Right Ascension (RA) centred at the position of the corresponding fiber axis. We estimate the typical fiber widths within these regions by fitting a single gaussian function to each of the above cuts (red dashed lines). As seen in the different subpanels, most of the observed fiber radial profiles can be well described by a Gaussian distribution with a full-width-half-maximum FWHM between 0.02 and 0.03~pc (cuts 1, 3, and 4).  
Asymmetric and complex profiles complicate this analysis producing a broader FWHM of $\sim$~0.05~pc (see cut 3).

\begin{figure*}
	\centering
	\includegraphics[width=\linewidth]{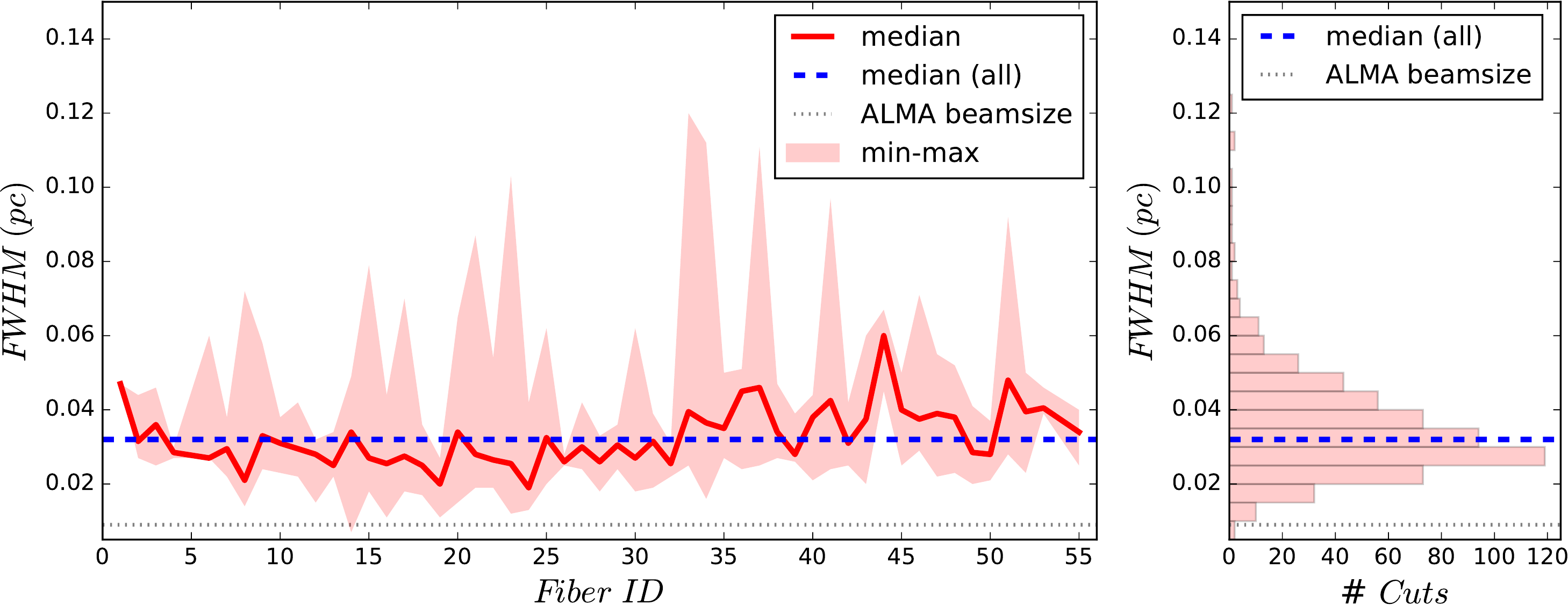}
	\caption{Observed fiber FWHM along the ISF. {\bf (Left)} Median (red solid line) and minimum-maximum (red shaded area) FWHM values obtained in each fiber. 
		 {\bf (Right)} Total FWHM values (576 data points; red histogram) measured along the ISF fibers.
		The total median value of 0.035~pc (blue dashed line) in comparison with the ALMA 0.009~pc beamsize (grey dotted line)  are indicated in both panels.
	}
	\label{fig:fiber_widths}
\end{figure*}

Similar to the above examples, we have statistically characterized the observed fiber widths using regular cuts perpendicular to the main fiber axes defined in Fig.~\ref{fig:ISF_ALMA}. 
We carried out this analysis in our total N$_2$H$^+$ integrated intensity maps (Fig.~\ref{fig:ISF_ALMA_SCUBA}, right panel).
Although potentially contaminated by fiber superpositions in some localized positions, 
this approach is preferred because the better stability of these integrated maps compared to the slightly noisier measurements derived from the line fits as well as for including the full intensity emission profile and not only high S/N data.	
Figure~\ref{fig:fiber_widths} (left) shows the median (solid red line) and total range (min-max; red shaded area) of the observed FWHM along all the ISF fibers (576 cuts). At each position, the reported FWHM values are obtained from the Gaussian fit of the total integrated N$_2$H$^+$ emission within the innermost 0.05~pc region around the fiber axis. The ISF fibers show  
a median FWHM of 0.035~pc (blue dashed line) with 85\% of the cuts showing FWHM values $<$~0.050~pc (see histogram in Fig.\ref{fig:fiber_widths} right). In contrast, less than 1\% of these cuts exhibit FWHM~$>$~0.1~pc. Systematic width variations of a factor $\sim$~2-3 are observed both between and within fibers. Larger widths are found in complex regions (e.g., fiber \#~23) or in structures with ill-defined axes (e.g., fiber \#~33). At the opposite end, several fibers are marginally resolved at the resolution of our ALMA observations, showing FWHM values of $\sim$~0.02~pc (e.g., fibers \#~15 and 24). 

Recent {\it Herschel} observations in nearby clouds like IC5146, Polaris, or Taurus
have suggested the existence of a constant filament width of $\sim$~0.1~pc \citep{ARZ11,AND14,PAL13}.
Independent studies have pointed out several observational biases affecting these measurements, questioning the robustness of these results \citep{SMI14,PAN17}.
Interferometric studies in clustered star-forming regions have also reported the detection of several elongated (fiber-like) substructures showing FWHMs below $<$~0.05~pc \citep{PIN11,FER14,HEN17}. 
The observed properties of the ISF fibers add new and direct evidence of a systematic departure from this ``universal'' behaviour. With widths ranging between $\sim$~0.02-0.05~pc, our statistical results undoubtedly prove the existence of filamentary structures with radial FWHM significantly narrower than the previously proposed 0.1~pc scale.

The compact widths of the ISF fibers could be related to the higher densities found in Orion when compared to other low-mass star-forming regions explored by {\it Herschel}.
Although highly idealized, it is instructive to compare the expected variations of the 
scale height of a filament in hydrostatic equilibrium ($H_0$) as function of the gas temperature (T$_K$), non-thermal motions ($\sigma_{NT}$), and density ($n$(H$_2$)) \citep{OST64}:
\begin{equation}\label{eq:density}
 H_0(T_\mathrm{K},n(\mathrm{H}_2))=\sqrt{\frac{2 \sigma_{eff}^2}{\pi\ G\  n(\mathrm{H}_2)}}\propto \sqrt{\frac{\mathrm{c}_{s}(T_\mathrm{K})^2+\sigma_{NT}^2}{n(\mathrm{H}_2)}}.
\end{equation}
Following this equilibrium solution, denser filaments are thus expected to show narrower radial configurations. 
For the ISF fibers, and consistent with our line opacity measurements (see Appendix~\ref{sec:fit_spectra:opacity}), we find typical gas densities of n(H$_2$)~$\sim10^7-10^8$~cm$^{-3}$ assuming a cylindrical symmetry and the previously derived M$_{\mathrm{lin}}$ and FWHM values (see Table~\ref{table:fiber_prop}). These estimates are (at least) 2 orders of magnitude higher than the densities measured for typical low-mass filaments in clouds like IC5146, Polaris, or Taurus.
According to Eq.~\ref{eq:density}, the observed radii in both the ISF fibers (T$_K\sim$~25~K; $\sigma_{NT}=\mathrm{c}_{s}$(25K);  n(H$_2$)~=~$10^8$~cm$^{-3}$) and {\it Herschel} filaments (e.g., {\it Herschel}: T$_K$~=10~K;  $\sigma_{NT}=\mathrm{c}_{s}$(10K); n(H$_2$)~=~$5\times10^5$~cm$^{-3}$) are expected to show a size dependency such that $ \frac{H_0(\mathrm{{\it Herschel}})}{H_0(\mathrm{ISF})}\sim9$. Although deviating from these simplified predictions, the observed width ratio $\frac{FWHM(\mathrm{{\it Herschel}})}{FWHM(\mathrm{ISF})}=\frac{0.1}{\sim0.03}\sim3-4$ indicates that fibers might present a wide range of intrinsic widths depending on the initial gas densities.

\subsection{Environmental effects: OMC-1 vs. OMC-2}\label{sec:environment}

%
\begin{table*}
	\caption{Fiber properties in molecular clouds}             
	\label{table:fiber_prop}      
	\centering                          
	\begin{tabular}{c c c c c c c}        
		\hline\hline                 
		& ISF$^{(1)}$ & OMC-1$^{(1)}$ & OMC-2$^{(1)}$ & NGC1333$^{(2)}$ & B213-L1495$^{(3,\star)}$ & Musca $^{(4,\#)}$ \\    
		\hline
		Region	& Orion	& Orion	& Orion & Perseus & Taurus & Musca \\
		Distance (pc) & 414	& 414	& 414 & 236 & 140 & 140 \\	
		Mass range	& High-mass & High-mass & Intermediate &  Intermediate/Low & Low & Low \\
		Mode	& Clustered & Clustered & Clustered & Clustered & Isolated & Isolated \\
		M$_{\mathrm{lin}}$(total) (M$_\sun$~pc$^{-1}$)	 & $\sim$500$^{(6,7)}$ & $\sim$500$^{(6,7)}$ & $\sim$500$^{(6,7)}$ & $\sim$~200 & $\sim$~50$^{(8)}$ & $\sim$~25$^{(5)}$ \\
		
		\hline                        
		Observations	& ALMA & ALMA & ALMA & IRAM~30m & FCRAO & APEX \\
		$\theta_{FWHM}\ (")$ & 4.5 & 4.5 & 4.5 & 30 & 60 & 28.5 \\ 
		Resolution (pc)		& 0.009 & 0.009 & 0.009 & 0.034 & 0.040 & 0.019 \\
		\hline
		\# Fertile fibers	&	55	&	28	&	27 & 14 & 7 & 1\\
		Size (pc) 		& 2.5$\times$0.48 & 1.0$\times$0.48  & 1.5$\times$0.48 & 2.0$\times$ 0.4  & 10.0$\times$1.0 & 6.5$\times$0.5 \\
		Area (pc$^2$)	& 1.2 & 0.48  & 0.72 & 0.8 & 10 &  3.25 \\
		$\Sigma$(fertile) (pc$^{-2}$) & 45.8 & 58.3 & 37.5 & 17.5 & 0.7 &  0.3\\
		\hline
		FWHM (pc) & $\sim$~0.02-0.05 & $\sim$~0.02-0.05 & $\sim$~0.02-0.05 & $\sim$~0.1 & $\sim$~0.1$^{(8)}$ & 0.07$^{(5)}$ \\
		$\left< \mathrm{Length}\right>$ (pc)	&	0.16$\pm$0.10	&	0.15$\pm$0.09	&	0.16$\pm$0.11 &	0.4$\pm$0.2 &	1.0$\pm$0.2 & 6.5\\
		$\left<M_{lin}\right>$ (M$_\sun$~pc$^{-1}$)	&	23$\pm$11	&	20$\pm$18	&	26$\pm$21 &	34$\pm$22 &	24$\pm$19 & 26 \\
		$\left<n(H_2)\right>$ (cm$^{-3}$)	&	$1.0\times10^8$	&	$0.8\times10^8$	&	$1.2\times10^8$ &	$\sim5\times10^5$ & $\sim10^5$ & $\sim10^4$ \\
		
		$\left<T_\mathrm{K}\right>$ (K)	&	24.7$\pm$6.7	&	28.9$\pm$7.1	&	20.4$\pm$1.5 &	$\sim$~10 &	$\sim$~10 & $\sim$~10 \\
		$\left<\sigma_{NT}/c_s(T_\mathrm{K})\right>$	&	0.77$\pm$0.25	&	0.89$\pm$0.28	&	0.65$\pm$0.13 &	1.0$\pm$0.3 &	1.1$\pm$0.3 & 0.7\\
		$\left<\nabla V_{LSR}|_{global}\right>$ (km~s$^{-1}$~pc$^{-1}$)	&	4.6$\pm$5.9	&	6.5$\pm$7.4	&	2.6$\pm$2.3 &	0.8$\pm$0.8 &	0.4$\pm$0.4 & 0.3 \\
		$\left<\nabla V_{LSR}|_{local}\right>$ (km~s$^{-1}$~pc$^{-1}$)	&	12.8$\pm$7.7	&	16.1$\pm$9.2	&	9.5$\pm$3.3 &	1.5$\pm$0.6 &	1.3$\pm$0.5 & $\sim$~1-2  \\
		\hline
	        
	\end{tabular}
	\tablefoot{The values listed in this table correspond to the mean and 1-$\sigma$ dispersion in each case. We notice that not all the above parameters can be described by a normal distributions so the use of a gaussian dispersion is only indicative of the observed varability.
	{\bf References}: (1) This work; (2) \citet{HAC17b}; (3) \citet{HAC13}; (4) \citet{HAC16}; (5) \citet{KAI16}; (6) \citet{BAL87}; (7) \citet{JOH99}; (8) \citet{PAL13}; ($\star$) Only fertile fibers detected in N$_2$H$^+$ are considered. ($\#$) Assumed as single fertile fiber \citep[see][for a discussion]{HAC16}.
	}
	
\end{table*}

So far, we have considered the entire ISF as a single star-forming region based on its continuity at large scales \citep[e.g.,][]{BAL87}.
In addition to this global analysis, the wide-field coverage of our ALMA mosaics allows us to investigate the potential impact of distinct feedback and dynamical effects on the ISF fibers. 
Indeed, both thermal and kinematic gas properties in the OMC-1 cloud are directly influenced by the ONC activity and its global gravitational collapse \citep{HAC17b} compared to the more pristine conditions expected for the OMC-2 region.  
In this section, we explore these environmental effects by studying the fiber properties in each OMC-1 and OMC-2 clouds independently.

In Figure~\ref{fig:fiber_histo}, we display the distribution of the 
fiber lengths (left), average l.o.s. velocity dispersions (centre), and mass per-unit-lengths (right) in both OMC-1 (blue) and OMC-2 regions (red). We also list the mean and 1-$\sigma$ dispersion values in Table~\ref{table:fiber_prop} (see col.~2 and 3).
Overall, we find no systematic differences between these properties along the ISF. On average, both OMC-1 and OMC-2 regions have fibers with statistically similar length, velocity dispersion, and mass per-unit-length. Minor variations are, however, apparent in Figures \ref{fig:fiber_histo}~(left) and \ref{fig:fiber_histo}~(centre). The OMC-1 fibers appear to  show larger velocity dispersions and gradients than those in OMC-2 (see additional parameters in Table~\ref{table:fiber_prop}). 
While appealing, these differences are found within the 1-$\sigma$ dispersion estimates and can only be considered as tentative. Despite these local variations, the observed ISF fibers are found to present roughly uniform properties regardless of their local environment.

Morphologically speaking, 
the ISF fibers are distributed in multiple hub-like associations \citep{MYE09} in both OMC-1 and OMC-2 regions.  As seen in Fig.~\ref{fig:ISF_ALMA}, most OMC-1 fibers are oriented radially converging towards the OMC-1 South region \citep[see ][]{ROD92,WIS98}.  Although partially recovered in our fiber analysis, similar fan-like arrangements of fibers can also be recognized at smaller scales towards the centres of OMC-2 FIR 4, OMC-2 FIR 6, and OMC-1 Ridge (Fig.~\ref{fig:ISF_regions}). These local properties are highlighted in comparison with the sparse distribution of fibers outside these regions (e.g., along OMC-2 FIR 1).
The above fiber arrangements seem to be created by gravitational focusing effects and to respond
the local variations of the cloud potential \citep{HAR07,VAZ17,KUT17}. We note that all of the above fiber hubs coincide with the positions of different (stars+gas) mass concentrations along the ISF. Moreover, we find an apparent correspondence between the depth of the potential and the number of objects in these fiber arrangements, with more massive hubs fed by an increasing number of fibers (e.g., see OMC-1 in comparison with OMC-2 FIR 4). In several cases, we also observe an increase of the gas motions along the fibers in the proximity of these hubs as expected in a gravitationally dominated gas velocity field. In the most prominent example, our high resolution observations confirm the correlation between the orientation and global velocity gradients of the OMC-1 fibers induced by their global infall towards the current centre of mass of this cloud located in the vicinity of the OMC-1 South proto-cluster \citep[see green cross in Fig.~\ref{fig:ISF_ALMA};][see also Appendix~\ref{sec:fit_spectra:kinematics}]{HAC17a}. While additional kinematic analysis are needed to confirm these results, gravity appears to be the driving mechanism in shaping the spatial distribution of fibers within this cloud.

This fiber organization is reminiscent of the competitive accretion scenario for the formation of high-mass stars \citep[][and references therein]{BON07}. Rather than stellar objects, however, the location of the potential well would be self-generated by initial concentration of fibers. In this sense, multiple centres of collapse can be created depending on the complexity of these fiber networks (e.g., OMC-1 ridge, OMC-2 FIR 6, or OMC-2 FIR) and evolve in time responding to changes on the local centre of mass \citep[e.g., from the ONC to OMC-1 South; see ][]{HAC17a}.

\section{Discussion}\label{sec:discussion}

\subsection{Fiber properties in low- and high-mass filaments}\label{sec:discussion:prop}
\begin{figure*}
	\centering
	\includegraphics[width=\linewidth]{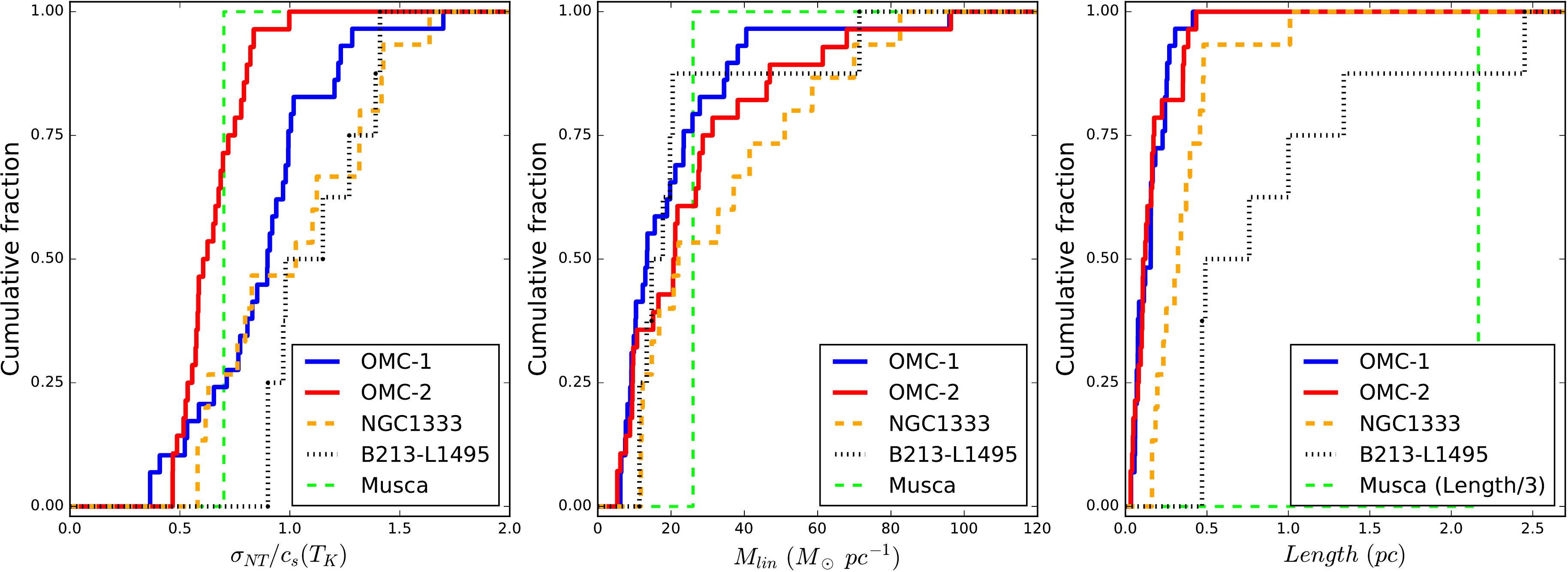}
	\caption{Statistical properties of the fertile fibers in Musca (green dashed line), B213-L1495 (black dotted line), NGC1333 (orange dashed line), OMC-2 (red solid line), and OMC-1 (blue solid line) (see also Table~\ref{table:fiber_prop}). Cumulative distributions of the line-of-sight, non-thermal velocity dispersion $\sigma_{NT}$ as a function of the local sound-speed c$_s(T_\mathrm{K})$ (left), mass-per-unit length (centre), and total fiber lengths (right).    
		 We notice that the total length of Musca has been reduced by a factor of 1/3 in the left panel.  
	}
	\label{fig:fiber_histo_clouds}
\end{figure*}


In combination with previous surveys, we systematically explore the variation of the fiber properties in both low- and high-mass filaments.  
We  statistically compare  these properties within a series of filamentary clouds with increasing total mass per-unit-length: (a) the isolated Musca filament \citep{HAC16}, (b) the low-mass L1495-B213 cloud in Taurus \citep{HAC13}, (c) the NGC~1333 ridge in Perseus \citep{HAC17b}, and both (d) the  OMC-2 plus (e) the OMC-1 regions in Orion (this work).
In Table~\ref{table:fiber_prop} we summarize the main characteristics of the star-forming fibers identified in these clouds. Typically detected in dense tracers like  N$_2$H$^+$, those fibers forming cores and stars are classified as fertile \citep[see ][for a discussion]{HAC13} 
\footnote{
	Our selection criteria assume as fertile all fibers with emission in the density selective  N$_2$H$^+$ line. That includes all the fibers identified in both the ISF (55) and NGC~1333 (14) regions, but only 7 out of the 35 fibers identified in B213-L1495 \citep[see][]{HAC13}. The other 28 fibers in B213-L1495, only detected at the lower densities traced by the C$^{18}$O (1-0) line, are considered sterile and are not included in our calculations (see Table~\ref{table:fiber_prop}).
	Although no N$_2$H$^+$ maps are available for Musca, we identify this filament as a single fertile fiber according to its fragmentation state showing multiple embedded prestellar cores and one protostar \citep{KAI16}.
}.
In Figure~\ref{fig:fiber_histo_clouds} we display the corresponding cumulative distributions for the observed l.o.s. non-thermal velocity dispersions (left), mass per-unit-length (centre), and fiber lengths (right) in all Musca (green dashed line), B213-L1495 (black dotted line), NGC~1333 (orange dashed line), OMC-2 (solid red line), and OMC-1 (solid blue line).

Across all regions, fibers are characterized by a low level of turbulence
relative to their local thermal motions. As seen in Fig.~\ref{fig:fiber_histo_clouds} (left), fibers exhibit l.o.s. non-thermal velocity dispersions ranging between $\sim$~0.5 and $\sim$~1.5 times their corresponding c$_s$(T$_\mathrm{K}$). We find no significant difference between fibers in clouds like Taurus, Perseus, and Orion, all showing similar average $\left< \sigma_{NT}/\mathrm{c}_s(T_\mathrm{K})\right>$ values within the 1-$\sigma$ error (see Table~\ref{table:fiber_prop}).
Their quiescent internal motions contrast with the broad linewidths commonly reported in these clouds. Our analysis indicates that most of these observational differences are created by a combination of global thermal effects, small-scale velocity gradients, and superposition, typically convolved in these previous single-dish surveys (see Sect.~\ref{sec:sigmaNT}). Once properly resolved, fibers show typical transonic motions (i.e., 1~$ < \sigma_{NT}/\mathrm{c}_s(T_\mathrm{K})\le$~2) in both low- and high-mass clouds. 

We also find a relatively good agreement between the distributions of the masses per-unit-length for fibers across the entire cloud spectrum.
Most of the fibers in Fig.~\ref{fig:fiber_histo_clouds} (centre) show consistent  M$_{\mathrm{lin}}$ values between 10 and $\sim$~80~M$_\sun$~pc$^{-1}$ in the four clouds used for this comparison. Small deviations can be attributed to low-number statistics (e.g., Musca or B213-L1495) and the limitations of our analysis (e.g., OMC-1 Ridge; see Appendix.~\ref{sec:HiFIVE}). 
In most cases, the reported M$_{\mathrm{lin}}$ values are consistent, within a factor of two, with the predicted masses for hydrostatic filaments internally supported by a combination of thermal and (sonic-like) turbulent motions (see Fig.~\ref{fig:stability}).
This apparent stable configuration might then explain the widespread detection of fibers over the wide range of conditions covered in these studies.

Contrary to the above similarities, we do find an environmental dependence on several of the observed fiber properties.
First, fibers in massive regions show larger internal velocity gradients than those in low-mass clouds. This increase is evidenced in the average local velocity gradients  $\left< \nabla V_{\mathrm{LSR|local}} \right>$ found in OMC-1 and OMC-2, typically between 3-5 times higher than those measured in Musca or Taurus (see Table~\ref{table:fiber_prop}). 
Part of these variations can be potentially affected by resolution effects and the increase of local motions at small scales \citep[e.g.][]{HAC11}.
Nevertheless, the largest $\left< \nabla V_\mathrm{LSR|local} \right>$ and $\left< \nabla V_\mathrm{LSR|global} \right>$ values in OMC-1 suggest that most of these differences would reflect the stronger influence of gravity affecting the gas bulk motions within clustered regions (see Sect.~\ref{sec:environment}). 

The observed fiber lengths in massive filaments also appear to be smaller compared to those found at low-mass regimes. 
In Fig.~\ref{fig:fiber_histo_clouds} (right), we identify a systematic shortening of the fiber lengths as function of the cloud mass. A two-sided Kolmogorov-Smirnov test of the observed fiber lengths in Taurus and Orion indicates a negligible probability ($<10^{-4}$~\%) for these distributions to be drawn from the same sample.
These differences are also apparent from the analysis of the fiber widths (see Sect.~\ref{sec:fiber_widths}). Our ALMA data show a significant reduction of the observed fiber widths in dense environments like Orion (FWHM$\sim$~0.035~pc) respect to those in low-mass clouds like Taurus or Musca (FWHM$\sim$~0.1~pc) (see Table~\ref{table:fiber_prop}). 
In combination with their transonic dynamical state, the proposed density dependency of these fiber properties (see Sect.~\ref{sec:fiber_widths}) favours a scenario in which fibers likely emerge as part of a self-regulated process controlled by the thermal and density structures of their local environments.

The initial gas densities of the fibers may also set the conditions for local gravitational fragmentation.
Along both  OMC-1 \citep{TEI16} and OMC-2 \citep{KAI17}, interferometric studies have identified typical Jeans-like separations between embedded sources (cores \& protostars) of $\lambda_{frag}=$~0.06-0.08~pc (12\,000-17\,000 AU).
These scales contrast with the typical inter-core distances in Taurus with $\lambda_{frag}\sim$~0.3 pc (60\,000 AU) \citep{TAF15}, that is, $\sim$~3.5-5 times larger than in Orion.
Although apparently different, the above variations are in relatively good agreement with the expected density dependence of the (critical) fragmentation scales for hydrostatic filaments, in which $\lambda_{crit}=3.94H$ or $\sim 2\ \times$~FWHM \citep{STO63}. 
In both Taurus and Orion, the observed fibers closely follow these predictions showing roughly similar $\frac{\lambda_\mathrm{frag}}{\mathrm{FWHM}}\sim 3$ \citep[see similar results in][]{SCH79}.

\subsection{Towards a unified model of star formation}\label{sec:discussion:unified}

The detection of fibers in Orion opens a new window on the description of the internal gas structure in massive clouds. Since their discovery \citep{HAC13}, different studies have reported the detection of fibers in isolated and clustered star-forming environments in clouds like \object{Perseus} \citep{HAC17b} and \object{Serpens}  \citep{LEE14,FER14}. 
Similar networks of small-scale filaments
are apparent in other nearby regions like \object{rho-Oph} \citep{AND07}, \object{B59} \citep{ROM09}, and \object{L1517} \citep{HAC11}.
Our new ALMA observations demonstrate that this underlying organization is not restricted to low-mass clouds but is also intrinsic of regions at higher mass regimes. Forming networks of different complexity, these fibers dominate the internal gas substructure independently of the environment, stellar content, or initial cloud conditions. In all cases, this pre-existing fiber substructure is directly linked to the position of stars and cores, likely controlling their formation.
The widespread detection of fibers seems to reflect the preferred organization mechanism for the dense molecular gas within clouds. While their origin remains under debate \citep{SMI14,MOE15,SMI16,CLA17}, the formation of these transonic fibers appears to be an inherent property of the ISM turbulence and an essential ingredient of the star formation process in both low- and high-mass regions.
 
\begin{figure}
	\centering
	\includegraphics[width=\linewidth]{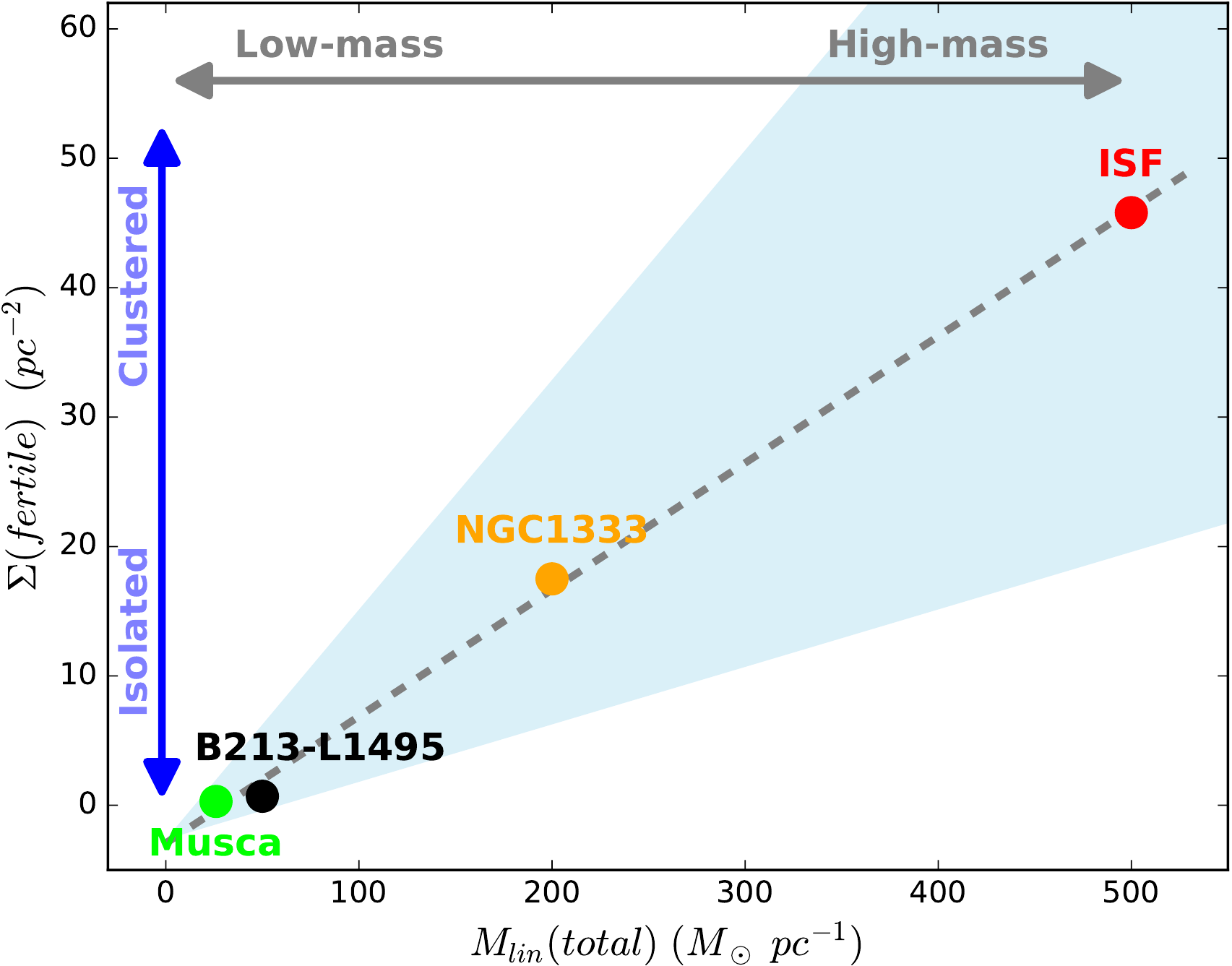}
	\caption{Empirical correlation between the total filament mass-per-unit-length (M$_{lin}(total)$) and the obtained surface density of fertile fibers ($\Sigma$(fertile)) in Musca (green), B213-L1495 (black), NGC1333 (orange), and ISF (red). The dashed line indicates the least-square fit of the above four clouds for which $\Sigma(\mathrm{fertile})\sim 0.1\times M_\mathrm{lin}\mathrm{(total)}$. The shaded blue area shows the expected variations of a factor of 2 respect to this linear correlation. 
	Blue and grey arrows indicate the proposed tendency of these fiber networks to form either isolated (bottom) or clustered (top) stellar populations and both low- (left) and high-mass (right) stars.
	}
	\label{fig:fiber_density}
\end{figure}

\citet{HAC17b} have proposed a correlation between the surface density of cores and protostars with the surface density of fertile fibers ($\Sigma$(fertile)). 
Under this hypothesis, a distributed population of stars would originate in clouds with an intrinsic low density of fertile fibers. On the other hand, more compact stellar systems would arise in densely populated fiber environments. 
Our observations extend these results towards massive clusters like the ONC. 
Originally postulated connecting low-mass regions like Taurus and Perseus, where $\frac{\Sigma(\mathrm{fertile,Per})}{\Sigma(\mathrm{fertile,Tau})} \sim 30$ \citep[see][]{HAC17b}, 
we find increasing values of $\frac{\Sigma(\mathrm{fertile,Ori})}{\Sigma(\mathrm{fertile,Tau})} \sim 50-80$ in comparison with Orion (see  in Table~\ref{table:fiber_prop}). 
As shown in Fig.~\ref{fig:fiber_density}, these reported ratios consistently increase with the typical mass-per-unit-length of these filamentary clouds (M$_{lin}(total)$) showing a surprising linear trend approximately described by a least-square fit $\Sigma(\mathrm{fertile})\sim 0.1\times M_\mathrm{lin}(\mathrm{total})$ (grey dashed line)\footnote{Local variations of a factor of $\sim$~2 in both  M$_\mathrm{lin}(\mathrm{total})$ and $\Sigma(\mathrm{fertile})$ values are apparent in all Taurus, Perseus, and Orion regions (e.g., see values for OMC-1 and OMC-2 in Table~\ref{table:fiber_prop}) introducing similar uncertainties in this linear fit (see blue shaded area in Fig.~\ref{fig:fiber_density}).}.
The lack of a complete census of protostars and cores in OMC-1 hampers a direct comparison between these fiber densities and the current young stellar population of this cloud \citep[see ][for a discussion]{HAC17b}. 
Nevertheless, the systematic increase of the fiber densities across two orders of magnitude suggests that the different star formation activity in these regions may be directly related to the intrinsic complexity of their fiber substructure.

The above variations of the fiber surface density could potentially unify our description of the star-formation process in  isolated and clustered regions.
In particular, Fig.~\ref{fig:fiber_density} indicates a continuity between the different star-formation scenarios classically distinguished in the literature (see arrows in this figure).
Rather than a distinct physical process (i.e., qualitative change), the origin of clusters in clouds like Orion  would be then explained as a scaled-up version (i.e., quantitative change) of those low-mass, Taurus-like regions forming stars in isolation. 
Moreover, the detection of the above fiber networks introduces an observational link between low-  and high-mass  star-formation mechanisms. 
Individually fragmenting and forming low-mass cores, these fibers would likely collide and merge in compact networks promoting the formation of localized high-mass, super-Jeans overdensities \citep[see also ][]{HAC17b}. Interestingly, this emergent behaviour would be naturally favoured by the mutual gravitational attraction between fibers in densely populated environments explaining the formation of massive objects at the centre of hub-like associations (Sect.~\ref{sec:environment}).

The above unified fiber scenario provides a direct prediction for future observations. 
According to Fig.~\ref{fig:fiber_density}, high-mass star-forming filaments with M$_\mathrm{lin}(\mathrm{total})\gg500$~M$_\sun$~pc$^{-1}$ are expected to be resolved into dense bundles of fibers in which $\Sigma(\mathrm{fertile})\gtrsim 50$~pc$^{-2}$.
Following our results in Orion (see \ref{sec:fiber_widths}), these compact configurations would be possible if fibers continue reducing their physical dimensions in higher density regimes.   
Current observational evidence indicates increasing levels of substructure at sub-parsec scales in massive IRDCs \citep[e.g.,][]{PER14,BEU15,HEN16}. Additional high-resolution studies are needed to confirm these results in different galactic environments in young proto-clusters of increasing mass and complexity.

\section{Conclusions}

In this Paper I, we have investigated the dense gas substructure of the massive Integral Shape Filament (ISF) in Orion using a new set of ALMA Cycle-3 observations.
We have created a high-dynamic range map of the N$_2$H$^+$ (1-0) emission in the ISF combining two 148-pointing mosaics with previous large-scale IRAM~30m single-dish observations.
From the detailed analysis of the gas kinematics, we have systematically characterized the internal organization and physical properties of the dense gas (n(H$_2$)~$>10^5$~cm$^{-3}$) within both OMC-1 and OMC-2 regions down to resolutions of 0.009~pc (or $\sim$~2000~AU). The results of this work are summarized as follows:

\begin{enumerate}
	\item We conservatively identify a total of 55 velocity-coherent fibers along both OMC-1 and OMC-2 clouds. Forming a braided network, these fibers explain the complex substructure of the massive ISF as well as its current distribution of cores and protostars (Sect.~\ref{sec:densegas}). Systematically observed in regions like Musca, Taurus, Perseus, and Orion, our results highlight the importance of fibers as the fundamental dense components of both low- and high-mass filaments.
	
	\item  Independently of the environment, the ISF fibers present transonic non-thermal velocity dispersions with respect to their local temperature (Sect.~\ref{sec:sigmaNT}). Most of these fibers are largely supported by a combination of thermal and sonic-like turbulent motions showing a (sub-)critical mass distribution (Sect.~\ref{sec:stability}). This quiescent dynamical state promotes the initial conditions for subsequent gravitational fragmentation. With analogous properties to those of low-mass regions, the detection of transonic fibers seems to reflect the preferred gas organization prior to the formation of stars.  
	
	\item The ISF fibers present a narrow radial distribution showing typical FWHM widths with a median value of 0.035~pc (Sect.~\ref{sec:fiber_widths}). This value is between $\sim$~2-5 times smaller than that found in regions like Perseus or Taurus. In addition to previous works, the statistical study of the ISF fibers demonstrates the existence of filamentary structures with compact radial distributions narrower than the so-called universal 0.1~pc width.
	
	\item Additionally, we observe a systematic shortening of the fibers between regions like Taurus and Orion (Sect.~\ref{sec:fiber_widths}). Both width and length variations seem to have their origin in the significantly higher gas densities found in massive filaments ($n$(H$_2$)~$>10^7$~cm$^{-3}$) when compared to low-mass regions (i.e., $n$(H$_2$)~$\sim10^5$~cm$^{-3}$). This density dependency indicates that the physical dimensions of the fibers may be self-regulated by the initial gas conditions. 
	
	\item The ISF fibers are spatially organized in distinct hub-like associations. We find a tentative correlation between the complexity and global kinematics of these fiber networks with the total mass distribution within both OMC-1 and OMC-2. These results suggest that fibers are internally oriented following the local gravitational potential within the ISF (Sect.~\ref{sec:environment}; see also Appendix~\ref{sec:fit_spectra:fits}).  
	
	\item Finally, we find a systematic increase of the surface density of star-forming (fertile) fibers as a function of the total mass per-unit-length in filamentary regions like Musca, B213-L1495, NGC1333, and the ISF (Sect.~\ref{sec:discussion:unified}). Based on this empirical correlation, we propose a unified star-formation scenario where the observational differences between both low- and high-mass clouds, as well as both isolated and clustered regimes, naturally emerge from their initial concentration of fibers.
	
\end{enumerate}

\begin{acknowledgements}
	The authors thank the anonymous referee for helping to improve the quality of the paper.
	AH thanks Alvaro Sanchez-Monje, Yanett Contreras, Daniel Harsono, and the ALLEGRO team for their support during the reduction of the ALMA data. AH also thanks Andreas Burkert and Ewine van Dishoeck for their insightful discussions and comments.
	This work is part of the research programme VENI with project number 639.041.644, which is (partly) financed by the Netherlands Organisation for Scientific Research (NWO). MT and AH thank the Spanish MINECO for support under grant AYA2016-79006-P.
	JG acknowledges funding by the Austrian Science Fund (FWF) under project number P 26718-N27.
	This paper makes use of the following ALMA data: ADS/JAO.ALMA\#2015.1.00669.S. 
	 PST acknowledges support from the STFC (grant number ST/M001296/1).
	ALMA is a partnership of ESO (representing its member states), NSF (USA) and NINS (Japan), together with NRC (Canada) and NSC and ASIAA (Taiwan) and KASI (Republic of Korea), in cooperation with the Republic of Chile. The Joint ALMA Observatory is operated by ESO, AUI/NRAO and NAOJ.
	Based on observations carried out with the IRAM~30m Telescope. IRAM is supported by INSU/CNRS (France), MPG (Germany) and IGN (Spain).
    This research made use of APLpy, an open-source plotting package for Python \citep{aplpy}.
    This research made use of Astropy, a community-developed core Python package for Astronomy \citep{Astropy}.
    	This paper made use of the TOPCAT software \citep{topcat}. 
\end{acknowledgements}

%

\begin{thebibliography}{}
	
		
	\bibitem[Andr{\'e} et al.(2007)]{AND07} Andr{\'e}, P., Belloche, A., Motte, F., \& Peretto, N.\ 2007, \aap, 472, 519 
	
	\bibitem[Andr{\'e} et al.(2010)]{AND10} Andr{\'e}, P., Men'shchikov, A., Bontemps, S., et al.\ 2010, \aap, 518, L102 
	
	\bibitem[Andr{\'e} et al.(2014)]{AND14} Andr{\'e}, P., Di Francesco, J., Ward-Thompson, D., et al.\ 2014, Protostars and Planets VI, 27 
	
	\bibitem[Andr{\'e} et al.(2016)]{AND16} Andr{\'e}, P., Rev{\'e}ret, V., K{\"o}nyves, V., et al.\ 2016, \aap, 592, A54
	
	\bibitem[Arzoumanian et al.(2011)]{ARZ11} Arzoumanian, D., Andr{\'e}, P., Didelon, P., et al.\ 2011, \aap, 529, L6 
	
	\bibitem[Arzoumanian et al.(2013)]{ARZ13} Arzoumanian, D., Andr{\'e}, P., Peretto, N., \& K{\"o}nyves, V.\ 2013, \aap, 553, A119 
	
	\bibitem[Astropy Collaboration et al.(2013)]{Astropy} Astropy Collaboration, Robitaille, T.~P., Tollerud, E.~J., et al.\ 2013, \aap, 558, A33
	
	\bibitem[Bally et al.(1987)]{BAL87} Bally, J., Langer, W.~D., Stark, A.~A., \& Wilson, R.~W.\ 1987, \apjl, 312, L45 
	
	\bibitem[Bally(2008)]{BAL08} Bally, J.\ 2008, Handbook of Star Forming Regions, Volume I, 4, 459 
	 
	\bibitem[Bally et al.(2011)]{BAL11} Bally, J., Cunningham, N.~J., Moeckel, N., et al.\ 2011, \apj, 727, 113
	
	\bibitem[Barnard et al.(1927)]{BAR27} Barnard, E.~E., Frost, E.~B., \& Calvert, M.~R.\ 1927, [Washington] Carnegie institution of Washington, 1927
	
	\bibitem[Beaumont et al.(2013)]{BEAU13} Beaumont, C.~N., Offner, S.~S.~R., Shetty, R., Glover, S.~C.~O., \& Goodman, A.~A.\ 2013, \apj, 777, 173 
		
	\bibitem[Bergin \& Tafalla(2007)]{BER07} Bergin, E.~A., \& Tafalla, M.\ 2007, \araa, 45, 339 
	
	\bibitem[Bern{\'e} et al.(2014)]{BER14} Bern{\'e}, O., Marcelino, N., \& Cernicharo, J.\ 2014, \apj, 795, 13
	
	\bibitem[Beuther et al.(2015)]{BEU15} Beuther, H., Ragan, S.~E., Johnston, K., et al.\ 2015, \aap, 584, A67 
	
	\bibitem[Bonnell et al.(2007)]{BON07} Bonnell, I.~A., Larson, R.~B., \& Zinnecker, H.\ 2007, Protostars and Planets V, 149 
	
	\bibitem[Buckle et al.(2012)]{BUC12} Buckle, J.~V., Davis, C.~J., Francesco, J.~D., et al.\ 2012, \mnras, 422, 521
	
	\bibitem[Busquet et al.(2013)]{BUS13} Busquet, G., Zhang, Q., Palau, A., et al.\ 2013, \apjl, 764, L26 
	
	\bibitem[Caselli et al.(2002)]{CAS02} Caselli, P., Benson, P.~J., Myers, P.~C., \& Tafalla, M.\ 2002, \apj, 572, 238
	
	\bibitem[Chini et al.(1997)]{CHI97} Chini, R., Reipurth, B., Ward-Thompson, D., et al.\ 1997, \apjl, 474, L135 
	
	\bibitem[Clarke et al.(2017)]{CLA17} Clarke, S.~D., Whitworth, A.~P., Duarte-Cabral, A., \& Hubber, D.~A.\ 2017, \mnras, 468, 2489 
	
	\bibitem[Dutrey et al.(1991)]{DUT91} Dutrey, A., Duvert, G., Castets, A., et al.\ 1991, \aap, 247, L9
	
	\bibitem[Feh{\'e}r et al.(2016)]{FER16} Feh{\'e}r, O., T{\'o}th, L.~V., Ward-Thompson, D., et al.\ 2016, \aap, 590, A75
	
	\bibitem[Fern{\'a}ndez-L{\'o}pez et al.(2014)]{FER14} Fern{\'a}ndez-L{\'o}pez, M., Arce, H.~G., Looney, L., et al.\ 2014, \apjl, 790, L19
	
	\bibitem[Forbrich et al.(2014)]{FOR14} Forbrich, J., {\"O}berg, K., Lada, C.~J., et al.\ 2014, \aap, 568, A27
	
	\bibitem[Forbrich et al.(2016)]{FOR16} Forbrich, J., Rivilla, V.~M., Menten, K.~M., et al.\ 2016, \apj, 822, 93
	
	\bibitem[Friesen et al.(2017)]{FRI17} Friesen, R.~K., Pineda, J.~E., Rosolowsky, E., et al.\ 2017, \apj, 843, 63 
	
	\bibitem[Furlan et al.(2016)]{FUR16} Furlan, E., Fischer, W.~J., Ali, B., et al.\ 2016, \apjs, 224, 5
	
	\bibitem[Galv{\'a}n-Madrid et al.(2010)]{GAL10} Galv{\'a}n-Madrid, R., Zhang, Q., Keto, E., et al.\ 2010, \apj, 725, 17 
	
	\bibitem[Genzel et al.(1982)]{GEN82} Genzel, R., Ho, P.~T.~P., Bieging, J., \& Downes, D.\ 1982, \apjl, 259, L103 
	
	\bibitem[Getman et al.(2005)]{GET05} Getman, K.~V., Flaccomio, E., Broos, P.~S., et al.\ 2005, \apjs, 160, 319
	
	\bibitem[Goddi et al.(2011)]{GOD11} Goddi, C., Greenhill, L.~J., Humphreys, E.~M.~L., Chandler, C.~J., \& Matthews, L.~D.\ 2011, \apjl, 739, L13
	
	\bibitem[Goicoechea et al.(2015)]{GOI15} Goicoechea, J.~R., Teyssier, D., Etxaluze, M., et al.\ 2015, \apj, 812, 75

	\bibitem[Grosso et al.(2005)]{GRO05} Grosso, N., Feigelson, E.~D., Getman, K.~V., et al.\ 2005, \apjs, 160, 530

	\bibitem[Hacar \& Tafalla(2011)]{HAC11} Hacar, A., \& Tafalla, M.\ 2011, \aap, 533, A34 
	
	\bibitem[Hacar et al.(2013)]{HAC13} Hacar, A., Tafalla, M., Kauffmann, J., \& Kov{\'a}cs, A.\ 2013, \aap, 554, A55
	
	\bibitem[Hacar et al.(2016)]{HAC16} Hacar, A., Kainulainen, J., Tafalla, M., Beuther, H., \& Alves, J.\ 2016, \aap, 587, A97 

	\bibitem[Hacar et al.(2017a)]{HAC17a} Hacar, A., Alves, J., Tafalla, M., \& Goicoechea, J.~R.\ 2017, \aap, 606, A123 
	
	\bibitem[Hacar et al.(2017b)]{HAC17b} Hacar, A., Tafalla, M., \& Alves, J.\ 2017, arXiv:1703.07029
	
	\bibitem[Hartmann(2002)]{HAR02} Hartmann, L.\ 2002, \apj, 578, 914
	
	\bibitem[Hartmann \& Burkert(2007)]{HAR07} Hartmann, L., \& Burkert, A.\ 2007, \apj, 654, 988
	
	\bibitem[Heigl et al.(2016)]{HEI16} Heigl, S., Burkert, A., \& Hacar, A.\ 2016, \mnras, 463, 4301 
	
	\bibitem[Hennemann et al.(2012)]{HEN12} Hennemann, M., Motte, F., Schneider, N., et al.\ 2012, \aap, 543, L3
	
	\bibitem[Henning et al.(2010)]{HEN10} Henning, T., Linz, H., Krause, O., et al.\ 2010, \aap, 518, L95 
	
	\bibitem[Henshaw et al.(2014)]{HEN14} Henshaw, J.~D., Caselli, P., Fontani, F., Jim{\'e}nez-Serra, I., \& Tan, J.~C.\ 2014, \mnras, 440, 2860 
	
	\bibitem[Henshaw et al.(2016)]{HEN16} Henshaw, J.~D., Caselli, P., Fontani, F., et al.\ 2016, \mnras, 463, 146 

	\bibitem[Henshaw et al.(2017)]{HEN17} Henshaw, J.~D., Jim{\'e}nez-Serra, I., Longmore, S.~N., et al.\ 2017, \mnras, 464, L31
	
	\bibitem[Hill et al.(2011)]{HIL11} Hill, T., Motte, F., Didelon, P., et al.\ 2011, \aap, 533, A94 
	
	\bibitem[Ikeda et al.(2007)]{IKE07} Ikeda, N., Sunada, K., \& Kitamura, Y.\ 2007, \apj, 665, 1194 
	
	\bibitem[Inutsuka \& Miyama(1997)]{INU97} Inutsuka, S.-i., \& Miyama, S.~M.\ 1997, \apj, 480, 681 
	
	\bibitem[Jackson et al.(2010)]{JAC10} Jackson, J.~M., Finn, S.~C., Chambers, E.~T., Rathborne, J.~M., \& Simon, R.\ 2010, \apjl, 719, L185
	
	\bibitem[Johnstone \& Bally(1999)]{JOH99} Johnstone, D., \& Bally, J.\ 1999, \apjl, 510, L49 
	
	\bibitem[Kainulainen et al.(2013)]{KAI13} Kainulainen, J., Ragan, S.~E., Henning, T., \& Stutz, A.\ 2013, \aap, 557, A120
	
	\bibitem[Kainulainen et al.(2016)]{KAI16} Kainulainen, J., Hacar, A., Alves, J., et al.\ 2016, \aap, 586, A27 

	\bibitem[Kainulainen et al.(2017)]{KAI17} Kainulainen, J., Stutz, A.~M., Stanke, T., et al.\ 2017, \aap, 600, A141
	
	\bibitem[Kauffmann et al.(2017)]{KAU17} Kauffmann, J., Goldsmith, P.~F., Melnick, G., et al.\ 2017, arXiv:1707.05352 
	
	\bibitem[Kounkel et al.(2014)]{KOU14} Kounkel, M., Hartmann, L., Loinard, L., et al.\ 2014, \apj, 790, 49 
	
	\bibitem[Kuznetsova et al.(2017)]{KUT17} Kuznetsova, A., Hartmann, L., \& Burkert, A.\ 2017, \apj, 836, 190
	
	\bibitem[Lee et al.(2014)]{LEE14} Lee, K.~I., Fern{\'a}ndez-L{\'o}pez, M., Storm, S., et al.\ 2014, \apj, 797, 76 
	
	\bibitem[Li et al.(2013)]{LI13} Li, D., Kauffmann, J., Zhang, Q., \& Chen, W.\ 2013, \apjl, 768, L5
	
	\bibitem[Li et al.(2016)]{LI16} Li, G.-X., Urquhart, J.~S., Leurini, S., et al.\ 2016, \aap, 591, A5
	
	\bibitem[Lombardi et al.(2014)]{LOM14} Lombardi, M., Bouy, H., Alves, J., \& Lada, C.~J.\ 2014, \aap, 566, A45 
	
	\bibitem[Martin-Pintado et al.(1990)]{MAR90} Martin-Pintado, J., Rodriguez-Franco, A., \& Bachiller, R.\ 1990, \apjl, 357, L49
	
	\bibitem[McMullin et al.(2007)]{CASA07} McMullin, J.~P., Waters, B., Schiebel, D., Young, W., \& Golap, K.\ 2007, Astronomical Data Analysis Software and Systems XVI, 376, 127 
	
	\bibitem[Meingast et al.(2016)]{MEI16} Meingast, S., Alves, J., Mardones, D., et al.\ 2016, \aap, 587, A153
	
	\bibitem[Megeath et al.(2012)]{MEG12} Megeath, S.~T., Gutermuth, R., Muzerolle, J., et al.\ 2012, \aj, 144, 192 
	
	\bibitem[Menten et al.(2007)]{MEN07} Menten, K.~M., Reid, M.~J., Forbrich, J., \& Brunthaler, A.\ 2007, \aap, 474, 515
	
	\bibitem[Mezger et al.(1990)]{MEZ90} Mezger, P.~G., Zylka, R., \& Wink, J.~E.\ 1990, \aap, 228, 95 
	
	\bibitem[Moeckel \& Burkert(2015)]{MOE15} Moeckel, N., \& Burkert, A.\ 2015, \apj, 807, 67
	
	\bibitem[Molinari et al.(2010)]{MOL10} Molinari, S., Swinyard, B., Bally, J., et al.\ 2010, \aap, 518, L100  
	
	\bibitem[Motte et al.(2017)]{MOT17} Motte, F., Bontemps, S., \& Louvet, F.\ 2017, arXiv:1706.00118
	
	\bibitem[Muench et al.(2008)]{MUN08} Muench, A., Getman, K., Hillenbrand, L., \& Preibisch, T.\ 2008, Handbook of Star Forming Regions, Volume I, 4, 483
	
	\bibitem[Myers \& Benson(1983)]{MYE83} Myers, P.~C., \& Benson, P.~J.\ 1983, \apj, 266, 309 

	\bibitem[Myers(2009)]{MYE09} Myers, P.~C.\ 2009, \apj, 700, 1609 

	\bibitem[O'dell(2001)]{ODE01} O'dell, C.~R.\ 2001, \araa, 39, 99
	
	\bibitem[O'Dell et al.(2008)]{ODE08} O'Dell, C.~R., Muench, A., Smith, N., \& Zapata, L.\ 2008, Handbook of Star Forming Regions, Volume I, 4, 544

	\bibitem[Ostriker(1964)]{OST64} Ostriker, J.\ 1964, \apj, 140, 1056 
	
	\bibitem[Pagani et al.(2009)]{PAG09} Pagani, L., Daniel, F., \& Dubernet, M.-L.\ 2009, \aap, 494, 719 
	
	\bibitem[Palau et al.(2017)]{PAL17} Palau, A., Zapata, L.~A., Roman-Zuniga, C.~G., et al.\ 2017, arXiv:1706.04623 
	
	\bibitem[Palmeirim et al.(2013)]{PAL13} Palmeirim, P., Andr{\'e}, P., Kirk, J., et al.\ 2013, \aap, 550, A38 
	
	\bibitem[Panopoulou et al.(2017)]{PAN17} Panopoulou, G.~V., Psaradaki, I., Skalidis, R., Tassis, K., \& Andrews, J.~J.\ 2017, \mnras, 466, 2529 
	
	\bibitem[Peretto \& Fuller(2009)]{PER09} Peretto, N., \& Fuller, G.~A.\ 2009, \aap, 505, 405
	
	\bibitem[Peretto et al.(2013)]{PER13} Peretto, N., Fuller, G.~A., Duarte-Cabral, A., et al.\ 2013, \aap, 555, A112 
	
	\bibitem[Peretto et al.(2014)]{PER14} Peretto, N., Fuller, G.~A., Andr{\'e}, P., et al.\ 2014, \aap, 561, A83 
	
	\bibitem[Peterson \& Megeath(2008)]{PET08} Peterson, D.~E., \& Megeath, S.~T.\ 2008, Handbook of Star Forming Regions, Volume I, 4, 590
	
	\bibitem[Pety et al.(2016)]{PET16} Pety, J., Guzm{\'a}n, V.~V., Orkisz, J.~H., et al.\ 2016, arXiv:1611.04037 
	
	\bibitem[Pineda et al.(2011)]{PIN11} Pineda, J.~E., Goodman, A.~A., Arce, H.~G., et al.\ 2011, \apjl, 739, L2 
	
	\bibitem[Rivilla et al.(2013)]{RIV13} Rivilla, V.~M., Mart{\'{\i}}n-Pintado, J., Sanz-Forcada, J., Jim{\'e}nez-Serra, I., \& Rodr{\'{\i}}guez-Franco, A.\ 2013, \mnras, 434, 2313 
	
	
	\bibitem[Robitaille \& Bressert(2012)]{aplpy} Robitaille, T., \& Bressert, E.\ 2012, Astrophysics Source Code Library, ascl:1208.017
	
	\bibitem[Rodriguez-Franco et al.(1992)]{ROD92} Rodriguez-Franco, A., Martin-Pintado, J., Gomez-Gonzalez, J., \& Planesas, P.\ 1992, \aap, 264, 592 
	
	\bibitem[Rom{\'a}n-Z{\'u}{\~n}iga et al.(2009)]{ROM09} Rom{\'a}n-Z{\'u}{\~n}iga, C.~G., Lada, C.~J., \& Alves, J.~F.\ 2009, \apj, 704, 183
	
	\bibitem[Salji et al.(2015)]{SAL15} Salji, C.~J., Richer, J.~S., Buckle, J.~V., et al.\ 2015, \mnras, 449, 1782 
	
	\bibitem[Schisano et al.(2014)]{SCH14} Schisano, E., Rygl, K.~L.~J., Molinari, S., et al.\ 2014, \apj, 791, 27
	
	\bibitem[Schneider \& Elmegreen(1979)]{SCH79} Schneider, S., \& Elmegreen, B.~G.\ 1979, \apjs, 41, 87
	
	\bibitem[Schneider et al.(2010)]{SCHN10} Schneider, N., Csengeri, T., Bontemps, S., et al.\ 2010, \aap, 520, A49
	
	\bibitem[Shimajiri et al.(2011)]{SHI11} Shimajiri, Y., Kawabe, R., Takakuwa, S., et al.\ 2011, \pasj, 63, 105 
	\bibitem[Shimajiri et al.(2014)]{SHI14} Shimajiri, Y., Kitamura, Y., Saito, M., et al.\ 2014, \aap, 564, A68
	
	\bibitem[Smith et al.(2014)]{SMI14} Smith, R.~J., Glover, S.~C.~O., \& Klessen, R.~S.\ 2014, \mnras, 445, 2900 
	
	\bibitem[Smith et al.(2016)]{SMI16} Smith, R.~J., Glover, S.~C.~O., Klessen, R.~S., \& Fuller, G.~A.\ 2016, \mnras, 455, 3640
	
	\bibitem[Stod{\'o}lkiewicz(1963)]{STO63} Stod{\'o}lkiewicz, J.~S.\ 1963, \actaa, 13, 30 
	
	\bibitem[Stutz et al.(2013)]{STU13} Stutz, A.~M., Tobin, J.~J., Stanke, T., et al.\ 2013, \apj, 767, 36
	
	\bibitem[Stutz \& Kainulainen(2015)]{STU15} Stutz, A.~M., \& Kainulainen, J.\ 2015, \aap, 577, L6
	
	\bibitem[Subrahmanyan et al.(2001)]{SUB01} Subrahmanyan, R., Goss, W.~M., \& Malin, D.~F.\ 2001, \aj, 121, 399 
	
	
	\bibitem[Tafalla \& Hacar(2015)]{TAF15} Tafalla, M., \& Hacar, A.\ 2015, \aap, 574, A104 
	
	\bibitem[Takahashi et al.(2013)]{TAK13} Takahashi, S., Ho, P.~T.~P., Teixeira, P.~S., Zapata, L.~A., \& Su, Y.-N.\ 2013, \apj, 763, 57
	
	\bibitem[Tatematsu et al.(2008)]{TAT08} Tatematsu, K., Kandori, R., Umemoto, T., \& Sekimoto, Y.\ 2008, \pasj, 60, 407
	
	\bibitem[Taylor(2005)]{topcat} Taylor, M.~B.\ 2005, Astronomical Data Analysis Software and Systems XIV, 347, 29 
	
	\bibitem[Teixeira et al.(2016)]{TEI16} Teixeira, P.~S., Takahashi, S., Zapata, L.~A., \& Ho, P.~T.~P.\ 2016, \aap, 587, A47
	
	\bibitem[van der Tak et al.(2007)]{VDT07} van der Tak, F.~F.~S., Black, J.~H., Sch{\"o}ier, F.~L., Jansen, D.~J., \& van Dishoeck, E.~F.\ 2007, \aap, 468, 627 
	
	\bibitem[V{\'a}zquez-Semadeni et al.(2017)]{VAZ17} V{\'a}zquez-Semadeni, E., Gonz{\'a}lez-Samaniego, A., \& Col{\'{\i}}n, P.\ 2017, \mnras, 467, 1313 
	
	\bibitem[Wiseman \& Ho(1998)]{WIS98} Wiseman, J.~J., \& Ho, P.~T.~P.\ 1998, \apj, 502, 676 
	
	
	\bibitem[Zamora-Avil{\'e}s et al.(2017)]{ZAM17} Zamora-Avil{\'e}s, M., Ballesteros-Paredes, J., \& Hartmann, L.~W.\ 2017, \mnras, 472, 647 
	

\end{thebibliography}
%

\begin{appendix}

\section{Fiber identification in high-dynamic range emission maps}\label{sec:HiFIVE}

As differentiating characteristic, fibers have been observationally defined as velocity-coherent structures identified by their spatial and velocity continuity. Exploiting this property in large molecular line datasets, \citet{HAC13} created the Friend-In-Velocity (FIVE) analysis technique. Compared to the direct analysis of the full 3D Position-Position-Velocity (PPV) analysis carried out by previous algorithms, FIVE combines a distinct (1+3D) treatment of the spectral data cubes. 
In a first stage, FIVE characterizes the individual kinematic properties of each gas component (e.g., centroid velocity V$_{lsr}$, line full-width-half-maximum $\Delta V$, peak intensity, and integrated area) extracted from the spectra (1D) using a supervised fitting technique.
After that, FIVE investigates the (3D) continuity of the previously extracted components from the analysis of their line centroids in the PPV space using a unique velocity gradient $\nabla V_{lsr,0}$ as linking parameter between nearby points. Connecting both velocity and spatial dimensions, this secondary process is carried out in two steps. FIVE firstly attempts to reconstruct the main skeletons of each individual structure from a selected subset of points with line intensities above a certain threshold I$_0$ surrounded by a given number N$_{friends}$ of PPV companions of similar properties. Once identified, FIVE uses these initial seeds to associate the rest of components following the same linking criteria \citep[see ][for additional details]{HAC13}.

FIVE has been successfully employed on the characterization of fibers in the B213-L1495 \citep[FIVE v1.0;][ $\theta_{FWHM}=$~0.040~pc]{HAC13} and NGC~1333 \citep[FIVE v1.1;][$\theta_{FWHM}=$~0.035~pc]{HAC17b} star-forming regions. In both cases, this FIVE analysis has benefited from the smooth velocity structure and the relatively small intensity variations observed within these clouds, allowing a direct identification of their fiber population using a single combination of [$\nabla V_{lsr,0}$,I$_0$,N$_{friends}$] values. Although multiple velocity coherent structures are clearly identified in our spectra, this simplified criterion is, however, challenged by the large intensity variations and the complex velocity structure found in Orion  (see also Appendix~\ref{sec:fit_spectra}). A series of individual tests indicate that no single combination of [$\nabla V_{lsr,0}$,I$_0$,N$_{friends}$] values is able to satisfactorily recover all the structures identified in our N$_2$H$^+$ integrated intensity maps in both OMC-1 and OMC-2 regions simultaneously (see Fig.~\ref{fig:ISF_ALMA_SCUBA}). 
The association in velocity carried out by FIVE is also hindered by the presence of highly variable velocity gradients at large-scales, dominating the kinematic structure of the OMC-1 region \citep[see][]{HAC17a}. Moreover, both intensity and velocity changes present a strong variability showing rapid changes at scales comparable to our ALMA beam size ($\theta_{FWHM}\sim$~0.009~pc), that is, $>$~4 times smaller than in any previous work. In the light of the above, a new approach is needed to investigate our high-dynamic range emission maps in Orion.

Motivated by the observed emission properties of the ISF (Sect.~\ref{sec:densegas}), we have created a new hierarchical version of FIVE, hereafter referred to as HiFIVE. This new HiFIVE analysis introduces three main variants to the original FIVE identification scheme optimizing the segmentation of hierarchically related substructures in clouds with large internal velocity variations. First, the initial line intensity $I_0$ threshold is replaced by a selection criterion based on the total gas column density $W_i$ (see the conversion between N$_2$H$^+$ intensities and total column densities in Sect.~\ref{sec:densegas}). Second, the former linking parameter $\nabla V_{lsr,0}$ is substituted by an adaptive velocity gradient $\nabla V_{lsr,i}=\frac{1}{2}\frac{\Delta V_i}{\theta_{FWHM}}$, self-defined from the local line Full-width-half-maximum $\Delta V_i$ and the beam size $\theta_{FWHM}$. Otherwise arbitrarily determined, this latter choice introduces a physically motivated definition of the linking parameter based on the resolution criterion by Nyquist in which two gaussian lines can only be distinguished if their centroids are separated by $\delta V_i\ge \Delta V_i/2$.
Third, and as new feature, HiFIVE carries out a hierarchical segmentation in three recursive levels $(W_i,N_i)$ (i=1,2,3) following a bottom-up approach. For that, the identification algorithm is first applied to the full dataset identifying the main kinematic structures (trees) recovered using a set of predefined $W_1$ and $N_1$ threshold. After that, a new algorithm execution is then independently carried out in each of these kinematic substructures identifying those separated fragments (branches) defined by a new selection criterion $W_2>W_1$ and $N_2\le N_1$. Finally, a third level of refinement (leaves) is then run for each of the previous new fragments with more than one independent substructure satisfying an intensity cut such that $W_3>W_2>W_1$ and $N_3\le N_2\le N_1$. The updates introduced in this new HiFIVE version reduce the selection parameters to the definition of the $(W_i,N_i)$ threshold pairs while the kinematic link between points is self-defined and locally adapted by the algorithm.

We have investigated the performance of HiFIVE with a reanalysis of the N$_2$H$^+$ emission in  NGC1333 \citep[see also][]{HAC17b}. Different tests indicate that the majority of the velocity-coherent fibers recovered by HIFIVE ($\sim$~70\%) are identified in the first stage of the algorithm (aka trees), while subsequent refinement levels (branches and leaves) help to disentangle some of the internal substructure in the most complex objects. As a rule of thumb, lower $(W_i,N_i)$ pairs lead into higher levels of fragmentation.  
The a priori arbitrary (and method-dependent) HiFIVE parameters suggests an statistical approach for the selection of the absolute $(W_i,N_i)$ thresholds. 
We thus adopted $W_1$, $W_2$, and $W_3$ values coincident with the $\sim$~50\% (median), 75\% (third quartile), and 90\% column densities values measured in N$_2$H$^+$ within this region and a similar number of friends $N_1=N_2=N_3=$~10. With this parameter choice, HiFIVE automatically recovers a total of 10 velocity-coherent structures within NGC1333, that is, 4 less than the 14 fibers identified by \citet{HAC17b} using a dedicated FIVE (v1.1) version. Differences in the total number of structures are created by the ambiguous assignment of components into fibers in regions with strong line multiplicity and compact velocity fields. Distinct merging and/or fragmentation effects are also observed by the different identification schemes used by these two algorithm versions. 
Similar conclusions are drawn from the reanalysis of the B213-L1495 region although the lower sensitivity of the spectra used in that study together with the inclusion of an additional tracers like C$^{18}$O hampers a more detailed comparison.
Overall, HiFIVE identifies a lower number of structures due to the larger linking gradients defined in regions showing lines with greater $\Delta V_i$ (see the connection between $\Delta V_i$ and $\nabla V_{lsr,i}$ above). 
Quantitative variations of $\sim40$~\% are thus expected in the estimated surface density of fibers $\Sigma$~(fertile) compared to the original FIVE. 
These limitations of the new HiFIVE version in complex regions are compensated by the enhanced adaptability of the new algorithm to the strong velocity variations in massive clouds (see Appendix~\ref{sec:fit_spectra:kinematics}).
Despite their differences, the mean fiber properties recovered in both HiFIVe and FIVE analysis (i.e., $\left< L\right>$, $\left< \sigma_{NT}/\mathrm{c}_s\right>$, $\left< \delta V_{LSR}\right>$, etc) appear to be consistent within 1-$\sigma$ level. 
While the individual assignment of components into different structures may vary depending on the algorithm version, our tests validate the statistical comparison of the new HiFIVE results with the analysis presented in previous studies.

\begin{figure*}[ht]
	\centering
	\includegraphics[width=1\linewidth]{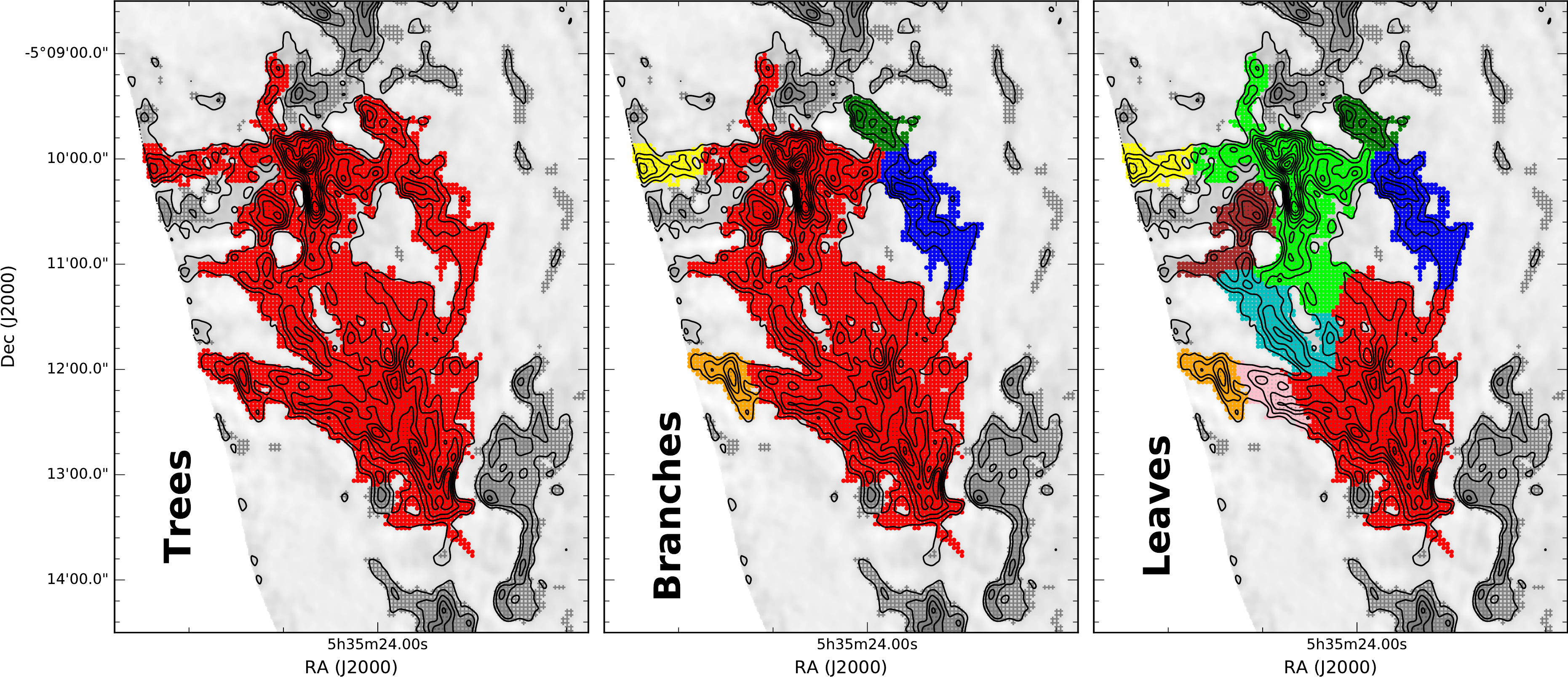}
	\caption{Structure identification in the OMC-2 region at different stages of the HiFIVE analysis. From left to right: (a) Trees, (b) Branches, and (c) Leaves (see text for a description). For simplicity, only those velocity structures extracted from the same initial tree are displayed in colours. The rest of the positions analysed by HiFIVE are indicated by grey crosses.  We notice that, although satisfactory in most cases, our HiFIVE analysis fails to describe some of the undoubtedly more complex gas substructure found in the ISF (e.g., fiber \#37, see central leave shown in red).
	}
	\label{fig:HIFIVE_performance}
\end{figure*}

Figure~\ref{fig:ISF_ALMA} (right) shows the results of the HiFIVE analysis carried our in the ISF.
Our analysis includes the (hyperfine) fit of $\sim$~70\ 000 ALMA N$_2$H$^+$ spectra from which we effectively extracted a total of $>$~25\ 000 individual components with S/N~$\ge$~3  (Appendix~\ref{sec:fit_spectra}). Similar to NGC1333, we define the internal structure of the ISF using three hierarchical levels determined by equivalent 
column density thresholds of $W_1=2.3\cdot10^{22}$~cm$^{-2}$ (or $\sim$~26 A$_v$; median), $W_2=4.0\cdot10^{22}$~cm$^{-2}$ (or $\sim$~44 A$_v$; third quartile), and $W_3=6.0\cdot10^{22}$~cm$^{-2}$ (or $\sim$~66 A$_v$; 90\% quantile) detected in our N$_2$H$^+$ spectra (see Sect.\ref{sec:densegas}), all with N$_i=$~10. As result of this analysis, HiFIVE identifies a total of 55 velocity-coherent substructures (fibers) along the ISF. The above parameter selection satisfactorily recovers most of the detected emission along this cloud assigning 95\% of the fitted components into fibers.

For visualization purposes, in Figure~\ref{fig:HIFIVE_performance} we illustrate the performance of the HiFIVE algorithm characterizing the gas substructure in the OMC-2 region. Overall,
we find an excellent correspondence between our HiFIVE velocity analysis and the structures identified in both  ALMA-N$_2$H$^+$ (Fig.~\ref{fig:ISF_ALMA}, right) and SCUBA maps (Fig.~\ref{fig:ISF_ALMA}, left).
Consistent with our previous tests, an inspection by eye of the above results indicate that HiFIVE fails to decompose several of the most complex structures within our maps. Examples of this behaviour can be found in the OMC-1 Ridge (fiber \#21), the OMC-2 FIR 6 cloud (fiber \#36), and the OMC-2 FIR 4 region (fiber \#43), all showing an additional complexity with multiple sub-branches not captured in our HiFIVE analysis (see Figure~\ref{fig:HIFIVE_performance}~c). Additional filamentary features not recovered HiFIVE are apparent at low intensities (e.g., see several fiber-like structures at the west side of the OMC-2 region running in parallel to the main ISF axis). 
Given the level of complexity observed in our N$_2$H$^+$ maps, the above 55 fibers should be interpreted as a simplified description of the real dense gas substructure within the ISF.

The reconstruction analysis carried out by our HiFIVE algorithm 
assumes the direct correspondence between the continuous structures identified in the observed Position-Position-Velocity (PPV) space and the real density structure of the Position-Position-Position (PPP) space.
Using simulations of clouds in global gravitational collapse, \citet{ZAM17} have recently suggested that some of the reconstructed fibers in the PPV space do not necessarily translate into physically coherent structures in PPP \citep[see also][for similar conclusions from a different suite of simulations]{MOE15}. 
Widely discussed in the past \citep[e.g., see][and references therein]{BEAU13}, \citet{ZAM17} remark how line-of-sight superpositions and projection effects can affect the observational identification of different gas structures in molecular clouds \citep[see also][]{SMI16}.
While potentially problematic in molecular line observations,
these contamination effects seem to have a limited impact in our ALMA data in Orion.
Those problematic cases found by Zamora-Avil\'es et al correspond to low-mass fibers identified in regions with complex superpositions.
Already in their simulations, however, the connection between both PPV and PPP spaces greatly improves in more massive and denser fibers. About two orders of magnitude larger than those found by \citet{ZAM17}, the detection of fibers in Orion at densities of $\gtrsim 10^7-10^8$~cm$^{-3}$ (Sect.~\ref{sec:densegas}) suggests a good correspondence with the mass distribution in the ISF. Reassuringly, most of the ISF fibers are also clearly separated in our emission maps (see Fig.~\ref{fig:ISF_ALMA}). 
With perhaps the exception of some dubious cases in our maps (see Appendix~\ref{sec:fit_spectra}), we therefore expect a close agreement between the fibers recovered by our HiFIVE analysis and the true gas substructure of the ISF.

\section{Line decomposition}\label{sec:fit_spectra}

\begin{figure*}
	\centering
	\includegraphics[width=0.8\linewidth]{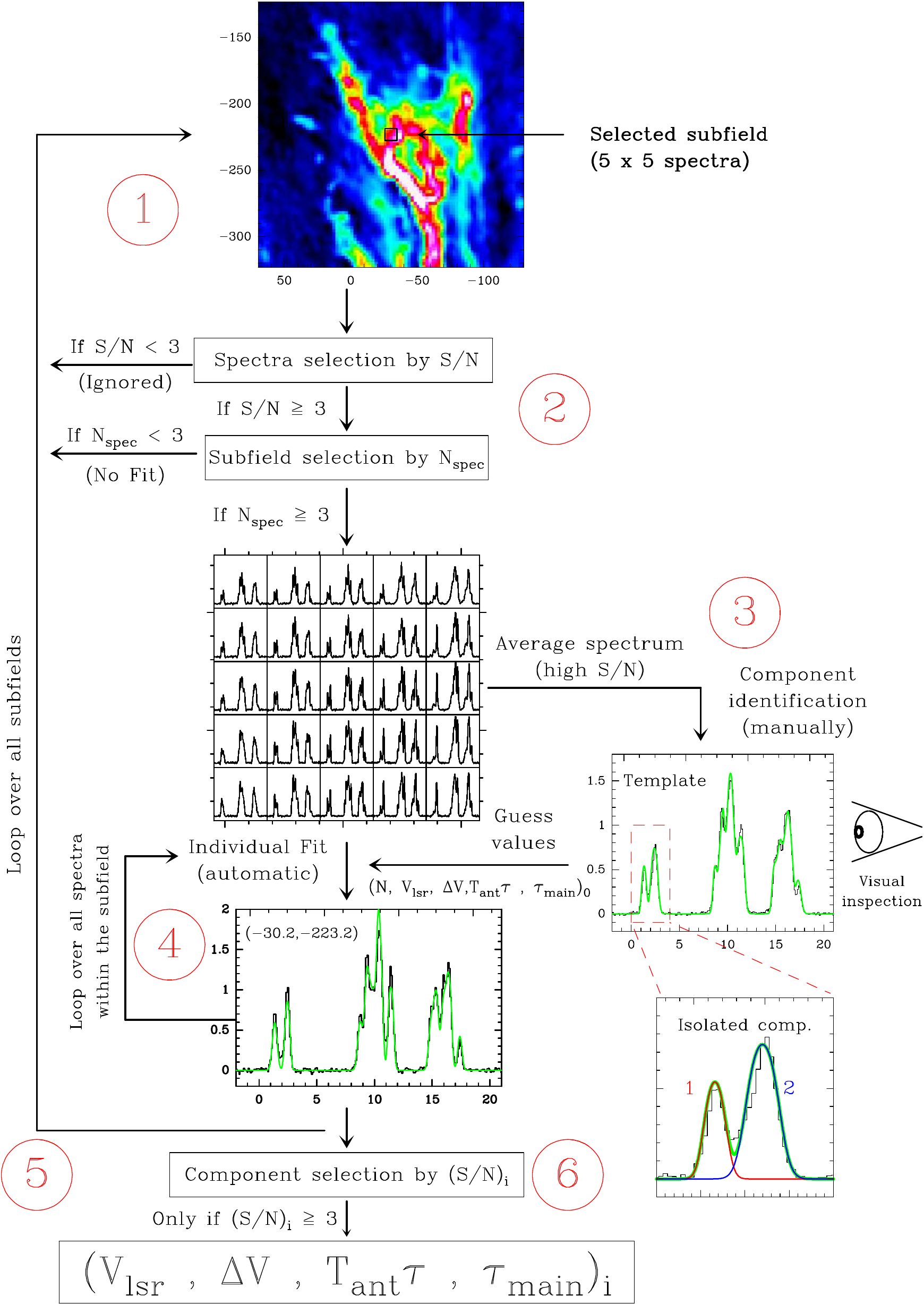}
	\caption{Workflow diagram describing the line fitting procedure of our N$_2$H$^+$ (1-0) ALMA data in Orion. The different steps are numbered in red. See text for a description.
	}
	\label{fig:fits_method}
\end{figure*}

The complexity of the gas velocity field observed in our Orion ALMA data requires a careful description of the line decomposition technique. In this Appendix, we describe the detailed line fitting procedure (Appendix~\ref{sec:fit_spectra:fits}) as well as the main line properties derived from this analysis (Appendix~\ref{sec:fit_spectra:kinematics} and \ref{sec:fit_spectra:opacity}).

\subsection{Fitting procedure}\label{sec:fit_spectra:fits}

Our spectral analysis adapts the fitting techniques introduced by \citet{HAC13} to the case of lines with hyperfine structure like N$_2$H$^+$.
Our fitting strategy combines both semi-automatic and supervised approaches for the analysis of large molecular datasets.
Figure~\ref{fig:fits_method} illustrates the workflow describing the full fitting procedure summarized in the following six steps:
\begin{enumerate}
	\item Subfield definition: The entire dataset is subdivided into subfields of 5$\times$5 spectra. The size of this subfield is estimated from the region in which the main properties of the individual spectra (number of components, mean velocity, etc) do not change significantly across the field.
	\item Selection: Within each subfield, a spectrum is only considered if it contains (at least) one emission channel above the mean rms in our dataset. A subfield is discarded if it does not contain at least three spectra fulfilling the above requirement.
	\item Subfield template: All accepted spectra are averaged to obtain a high S/N spectrum meant to describe the mean line properties within the subfield. This spectrum template is visually inspected and manually fitted. The number of components and the goodness of the fit are evaluated from the direct inspection of the isolated, optically thin $(JF_1F)=(101-012)$ hyperfine component. If necessary, individual spectra are examined in order to refine this fit. 
	\item Automatic fit: The fitting results obtained in the previous step are used as guess values for the automatic fit of all accepted spectra within the subfield. Although informed in their initial values, all fitting parameters are left unconstrained. 
	\item Iteration: Steps 1-4 are repeated over all the subfields in our maps.
	\item Component selection: A final S/N selection is carried out for every individual component with respect to the rms level in each spectrum. Only velocity components with individual peak intensities with S/N~$\ge$~3 in their central hyperfine $(JF_1F)=(123-012)$ line are accepted for their analysis.
\end{enumerate}
The entire fitting procedure is carried out with the GILDAS/CLASS software\footnote{http://www.iram.fr/IRAMFR/GILDAS}. All hyperfine lines are simultaneously fitted using the HFS method assuming similar excitation conditions and widths, as well as Gaussian opacity profiles for all components in the multiplet (i.e. steps 3 \& 4). If resolved in the isolated $(JF_1F)=(101-012)$ hyperfine component, the presence of multiple peaks in our spectra, are assumed to correspond to independent gas structures and are, therefore, fitted independently (see an example in Fig.~\ref{fig:fits_method}). In case of doubt, the spectra are conservatively fitted with the lowest possible number of components. Non-Gaussian components producing self-absorbed profiles or line wings can potentially contaminate this analysis. Although these effects are identified in some localized spectra, the visual inspection of the data indicates a statistically small influence in our large dataset.

\begin{figure*}
	\centering
	\includegraphics[width=\linewidth]{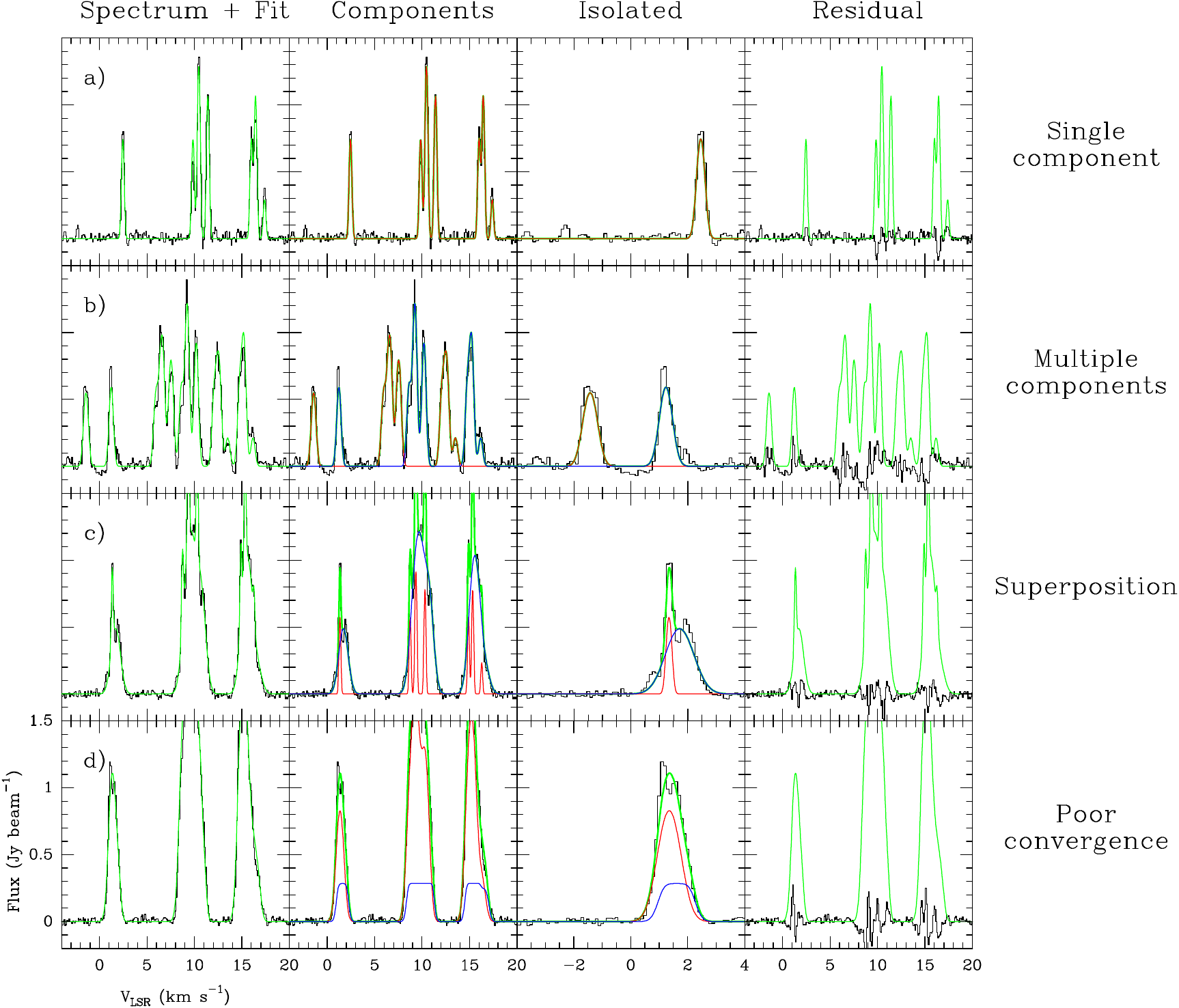}
	\caption{Examples of the N$_2$H$^+$ (1-0) ALMA spectra found in Orion: 
		From left to right: (1) observed spectrum and best-fit solution: (2) individual components identified by the fit; (3) zoom-in on the $(JF_1F)=(101-012)$ hyperfine component; (4) residual spectrum after the subtraction of the best-fit solution. For comparison, the best-fit solution (either single or multiple) is displayed in all subpanels in green. Each individual velocity component is denoted in red and blue in panels 2 and 3.
		From top to bottom: (a) single component; (b) well-resolved multiple components; (c) superposition of components; (d) problematic spectra. 
	}
	\label{fig:fits_example}
\end{figure*}

In Figure~\ref{fig:fits_example} we display a series of representative spectra fitted in our ALMA data. This figure includes both the total (column 1)  plus individual line profiles (cols. 2 and 3) as well as their residual emission after the subtraction of their best-fit solution (col. 4). Panels a and b shows the most common situations found in our spectra with either single or multiple well-resolved components with velocity differences larger than the individual linewidth, that is, $\delta V_{lsr}>\Delta V$. In these simple cases, small residuals (few percent of the total integrated emission) are created by the slightly different excitation temperatures of each independent hyperfine component not considered in our fits. Panel c illustrates a close superposition of components denoting the 
limiting case where two superposed lines can be resolved if $\delta V_{lsr}\gtrsim\Delta V/2$.
For comparison, Panel d illustrates an example of the problematic cases found in regions with extreme variability in velocity (i.e., not well described by their subfield template spectrum), unresolved multiplicity (e.g., too close to be separated in all spectra within the subfield), and/or with complex line profiles leading to poor or dubious fitting solutions (see individual components in this panel). Most of these last pathological effects are minimized by the conservative selection of components carried out in our analysis. Estimated to affect $<$~5\% of our fits, their results are included in our kinematic analysis in order to preserve the total line intensity\footnote{We note that fits similar to the one shown in panel d (with $\delta V_{lsr}\lesssim\Delta V/2$) are likely recovered by HiFIVE into a single structure in velocity (see Appendix~\ref{sec:HiFIVE}). With little impact on the derived kinematic properties of fibers the addition of these components allows us to recover the complete line emission.}.

Using the above technique, we analysed more than 70\ 000 N$_2$H$^+$ (1-0) spectra in Orion. A total of $>$~25\ 000 individual components are detected with S/N~$\ge$~3. For each of these components, the HFS method provides with the best-solution estimates for the line central velocity $V_{LSR}$, linewidth $\Delta V$, the intensity product $T_{ant}\tau$, and the total multiplet opacity $\tau_{main}$ from which all line properties can be derived (see CLASS cookbook; https://www.iram.fr/IRAMFR/GILDAS/doc/pdf/class.pdf). 
Among these quantities, both $V_{LSR}$ and $\Delta V$ are estimated within typical errors of 0.01~km~s$^{-1}$. With significantly less accuracy, median relative uncertainties of $\sim$~25\% are measured in our $\tau_{main}$ estimates.

\subsection{Fitting results (I): kinematics}\label{sec:fit_spectra:kinematics}

\begin{figure*}
	\centering
	\includegraphics[width=\linewidth]{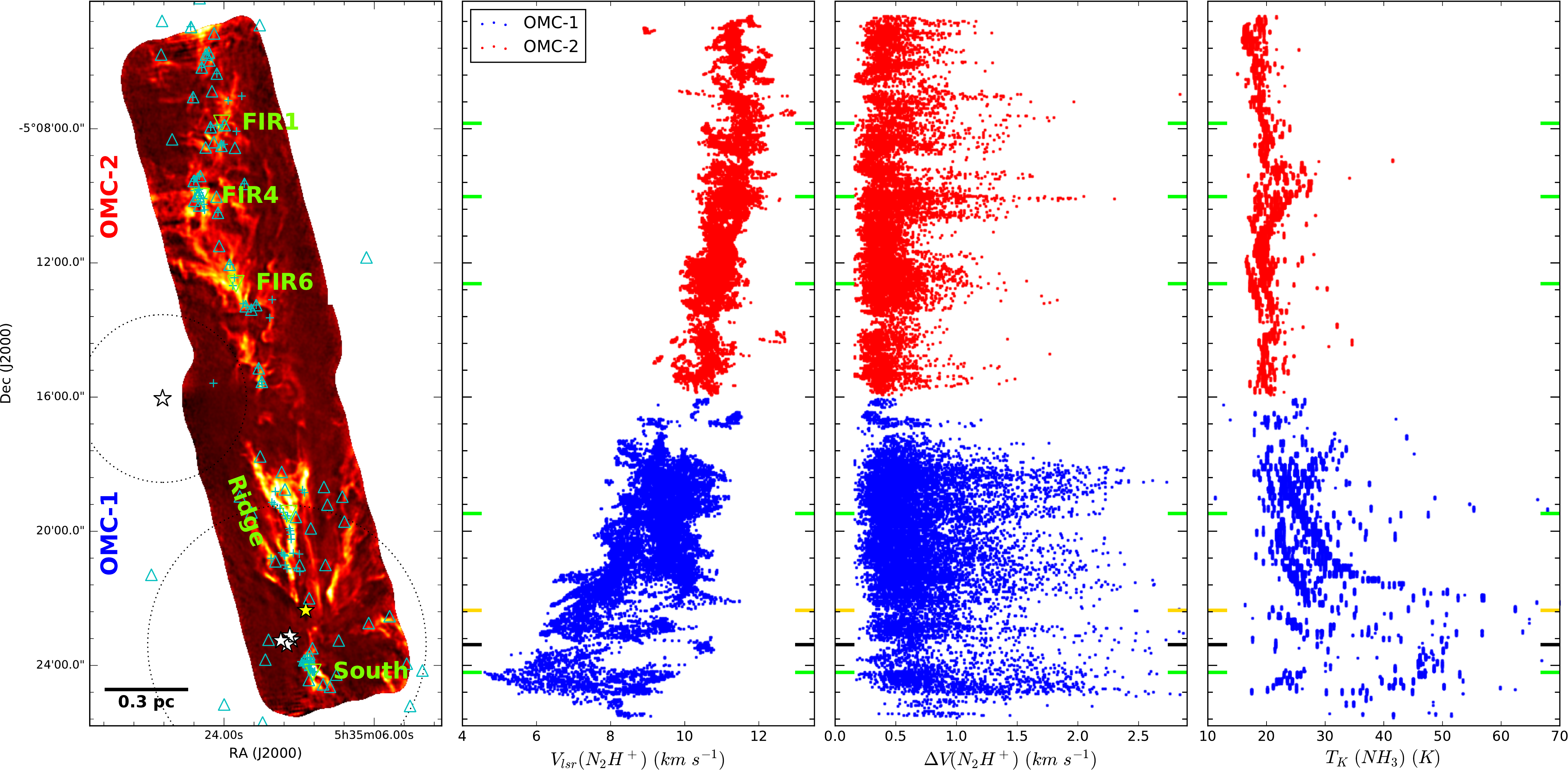}
	\caption{Gas kinematic properties as a function of declination along both OMC-1 (blue) and OMC-2 (red) regions. From left to right: (1) Integrated N$_2$H$^+$ (1-0) emission. Symbols are similar to those in Fig.~\ref{fig:ISF_ALMA_SCUBA}; (2) Centroid velocities and (3) linewidths for all the N$_2$H$^+$ (1-0) components detected in our ALMA observations with S/N~$\ge$~3; (4) NH$_3$-derived gas kinetic temperatures \citep{FRI17} at the position of our N$_2$H$^+$ measurements. For reference, the declination of the Trapezium stars (black), the Orion BN/KL region (yellow), and FIR sources (green) are indicated in all subplots.
	}
	\label{fig:fits_PV}
\end{figure*}

In Figure~\ref{fig:fits_PV} (first and second panels), we display the centroid velocities $V_{LSR}(N_2H^+)$ of all the components fitted in our N$_2$H$^+$ (1-0) spectra with S/N~$\ge$~3 as function of declination along the ISF. Overall, the gas velocity structure reproduces the results obtained by previous single-dish observations. Along the $\sim$~1.5~pc of the OMC-2 region, the average gas velocity is dominated by a smooth north-south velocity gradient of $\sim$~1~km~s$^{-1}$~pc$^{-1}$ \citep[e.g.,][]{BAL87}. Closer to the ONC, the observed gas velocities define a (blue-shifted) V-shape profile consistent the gravitational collapse of the OMC-1 region \citep{HAC17b}. These gravitationally induced motions generate the largest velocity differences along the ISF with maximum velocity shifts of $\sim$~5~km~s$^{-1}$ in the surroundings of OMC-1 South. Although largely dominated by the stellar component, the bottom of the potential locates the current centre of mass of the OMC-1 system (i.e. gas + stars) $\sim$~60" away from the Trapezium.
An additional kinematic complexity is revealed by ALMA at subparsec resolutions.
Organized velocity fluctuations are found in the proximity of OMC-1 Ridge, OMC-2 FIR-4, and OMC-2 FIR-6. Superposed on them, a large variety of velocity gradients, with intensities between 1-20~km~s$^{-1}$~pc$^{-1}$, are seen at scales of $\sim$~0.1-0.3~pc. In more detail, several velocity-coherent structures can be directly identified as groups of continuous velocity in this plot. 

Similar to the gas velocities, in Fig.~\ref{fig:fits_PV} (third panel) we illustrate the distribution of the N$_2$H$^+$ linewidths ($\Delta V(N_2H^+)$) in both OMC-1 and OMC-2 clouds. Across the entire ISF, 75\% of the observed linewidths show $\Delta V (N_2H^+)<$~0.71~km~s$^{-1}$.
We identify a systematic increase of the observed linewidths between OMC-2 ($\left< \Delta V (N_2H^+) \right> =$~0.48~km~s$^{-1}$) and OMC-1 ($ \left< \Delta V (N_2H^+)\right> =$~0.72~km~s$^{-1}$) partially correlated with the increasing gas kinetic temperatures measured in the proximity of the ONC (see also forth panel). Localized line broadening effects are also found at the position of most of the embedded FIR sources (e.g., OMC-2 FIR-4) and several of the embedded protostars within both OMC-1 and OMC-2 regions (see spikes in the $\Delta V$ distributions).

The reported line emission properties within the ISF
highlight  some of the challenges investigating the internal velocity field of massive filaments.
The characterization of these lines must consider the strong spatial variability observed in both velocity centroids and linewidths at the current ALMA resolution.
Moreover, their analysis needs to evaluate the combined influence of all global and local motions in combination with feedback effects. These findings have motivated the development of our new algorithm HiFIVE (see Appendix~\ref{sec:HiFIVE}).

\subsection{Fitting results (II): line opacities}\label{sec:fit_spectra:opacity}

In addition to the line kinematics, our fits provide detailed information of the total N$_2$H$^+$ (1-0) opacity ($\tau_{main}$) as well as the corresponding opacities for each of its individual hyperfine components (with $\tau_i = r_i\cdot\tau_{main}$ for an hyperfine line of relative intensity $r_i$). 
We have evaluated the opacity of the extracted N$_2$H$^+$ (1-0) fits from the analysis of 
the central $(JF_1F)=(123-012)$ transition $\tau(123-012)$, that is, the brightest and  most optically thick hyperfine line in the multiplet ($r_i=7/27$) and the one subject of largest saturation effects.
Among all the extracted components, $\sim$~75\% show central line opacities with $\tau(123-012)<$~1 (optically thin) and 92\% $\tau(123-012)<$~2 (moderate opacity). The other 8\% include both high opacity spectra as well as problematic fits (see Appendix~\ref{sec:fit_spectra:fits}). Most of these optically thick components are located within unresolved regions like OMC-1 Ridge, OMC-2 FIR~6 and OMC-2 FIR~4. In comparison, 99\% of the isolated hyperfine $(JF_1F)=(101-012)$ lines ($r_i=1/27$) are estimated to be optically thin. 

\begin{figure}[h]
	\centering
	\includegraphics[width=\linewidth]{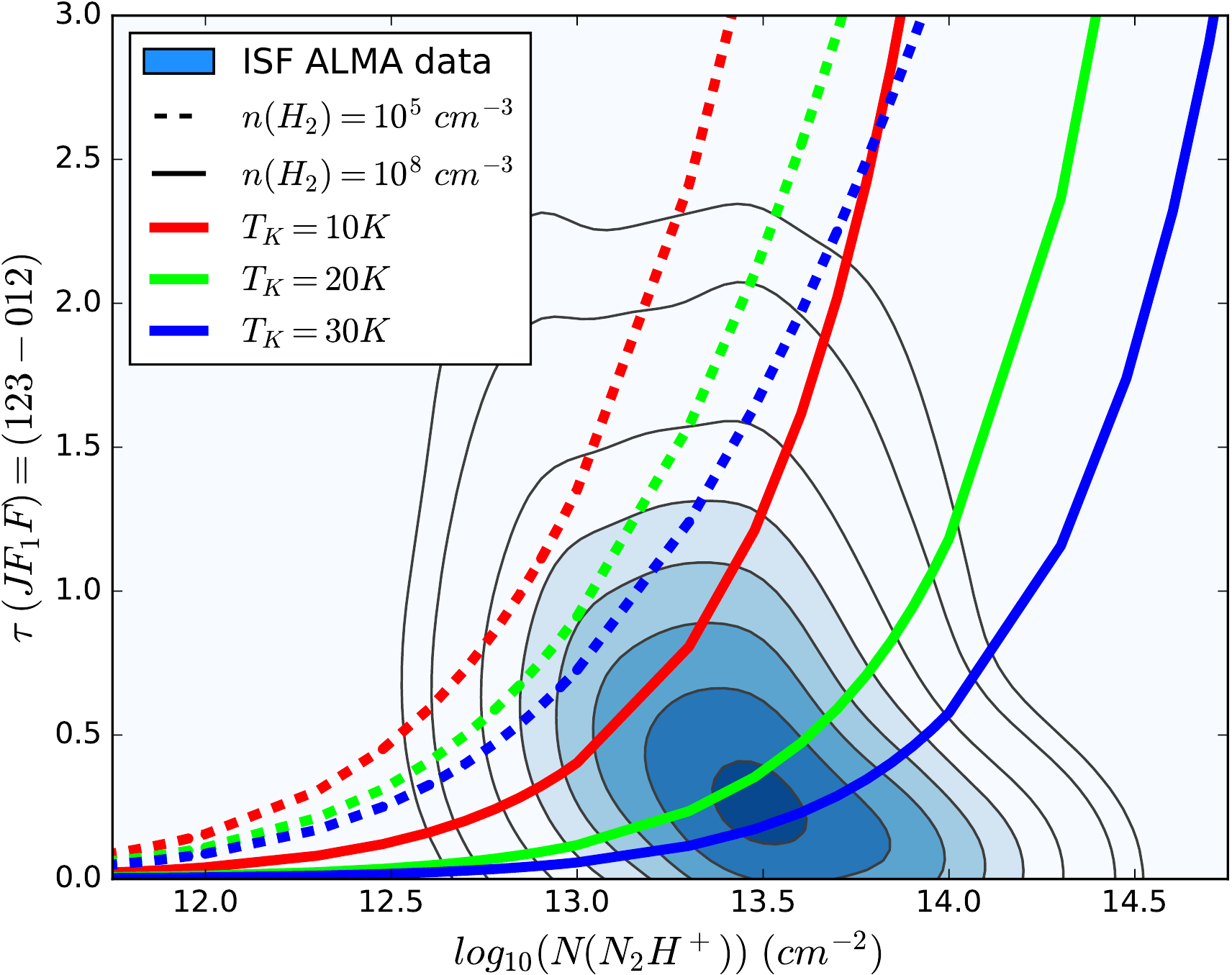}
	\caption{
		Distribution of the derived $\tau (JF_1F)=(123-012)$ opacities as function of the estimated total N$_2$H$^+$ column density in the ISF. The plot represents the kernel density estimates of $>$~25\,000 individual components extracted from our ALMA observations. The last contour encloses 95\% of the components fitted in our spectra. For comparison, we overplot the different RADEX predictions for an isothermal gas at kinetic temperatures of 10~K (red), 20~K (green), and 30~K (blue) at densities n(H$_2$) between 10$^5$~cm$^{-3}$ (dashed lines) and 10$^8$~cm$^{-3}$ (solid lines), all calculated assuming a constant linewidth of $\Delta V=0.7$~km~s$^{-1}$.}
	\label{fig:fits_opacities}
\end{figure}

In Figure~\ref{fig:fits_opacities} we have investigated the correlation between the observed $(JF_1F)=(123-012)$ line opacities and the total N$_2$H$^+$ column densities ($N(N_2H^+)$) in the ISF. We have estimated these latter $N(N_2H^+)$ values from the individual $\tau_{main}$ and $T_{ant}\tau$ line parameters following the standard approach introduced by \citet{CAS02}. Using a kernel density estimate, this figure summarizes the results of more than 25\ 000 components measured in our ALMA observations. We have compared our estimates with the radiative transfer predictions provided by RADEX \citep{VDT07}. On average, the observed N$_2$H$^+$ optically thin lines are consistent with the expected emission properties for the gas in the ISF (Sect.\ref{sec:densegas}). In particular, the relatively low opacities detected in our  N$_2$H$^+$ (1-0) spectra can be explained by the higher temperatures (20-30~K) and densities ($\sim 10^7-10^8$~cm$^{-3}$) found in Orion compared to the typical gas conditions in low-mass regions like Taurus (see models with 10~K and  $10^5$~cm$^{-3}$) leading to the depopulation of the J=1 level in favour of those at higher energies (J~$\ge$~3). 

High line opacities can crucially affect the mass estimates derived from the integrated emission of tracers like N$_2$H$^+$. Opacity effects can hide large column densities in the case of heavily thick lines. Our statistical analysis demonstrates that these effects are minimized in the case of Orion. Based on these results, we adopted an optically thin approximation for all our mass estimates (Sect.~\ref{sec:densegas}). Strictly speaking, this assumption is valid in 75\% of our measurements. Additional tests indicate a maximum (and selective) increase of $\sim$~40\% on the total gas masses including opacity effects (e.g., Sect.~\ref{sec:stability}). The large fitting uncertainties in heavily opaque lines prevent a more detailed treatment of these opacity effects. Still, their expected variations are within the uncertainties already associated to our mass conversion factor estimated of $\sim$~2-3 (see Eq.~\ref{eq:colden}) and are not considered in our analysis.

\section{Single-dish N$_2$H$^+$ abundances}\label{sec:abundance}

\begin{figure}
	\centering
	\includegraphics[width=\linewidth]{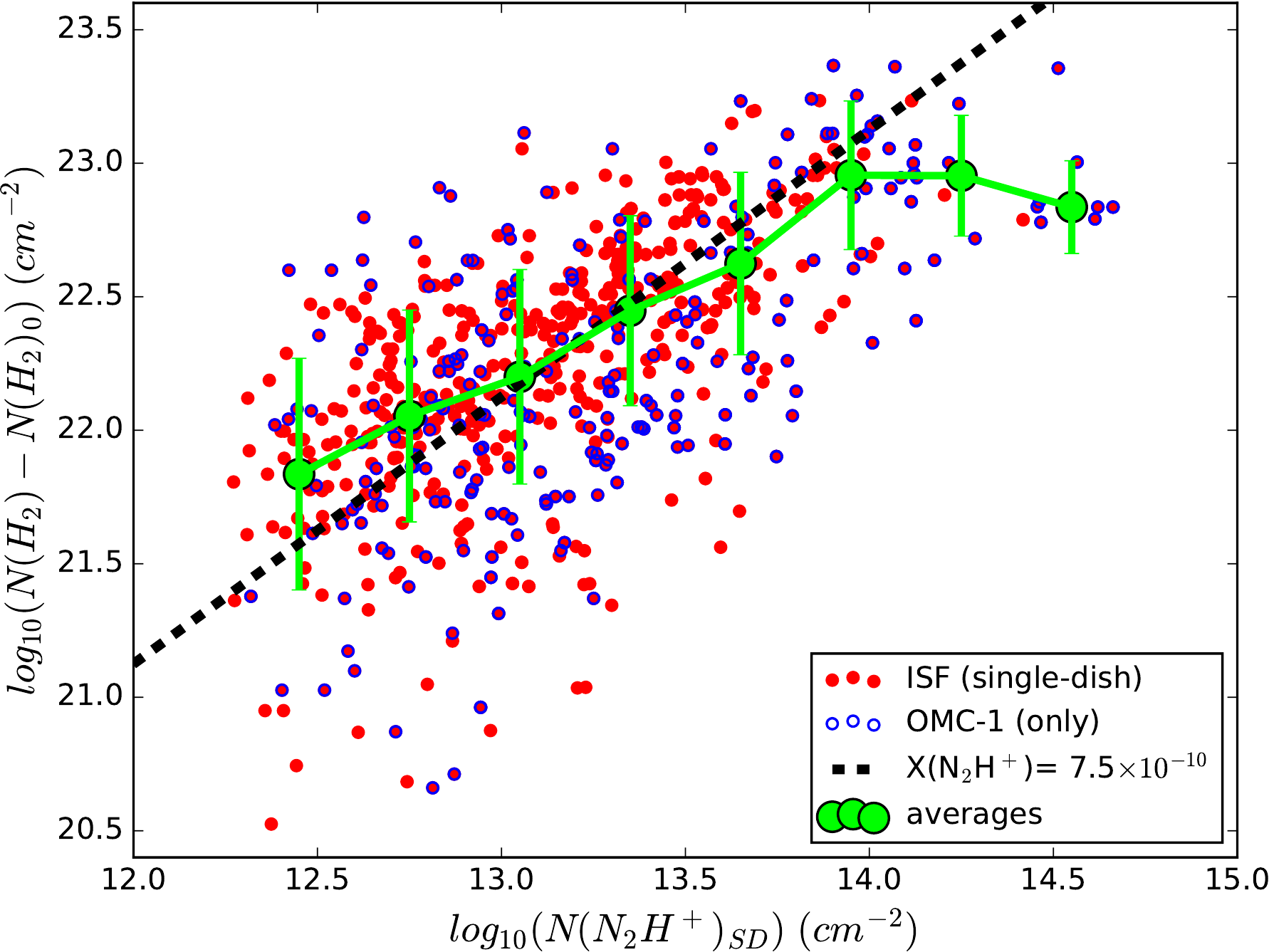}
	\caption{
		Derived total N$_2$H$^+$ column density in our single-dish observations ($N(N_2H^+)_{SD}$) as function of the (normalized) {\it Herschel} H$_2$ column density ($N(H_2)-N(H_2)_0$) for all the positions shown in Fig.~\ref{fig:N2H+_cal} along the ISF (red) (see text for further details). The positions in OMC-1 are highlighted in blue. Both mean (points) and 1-$\sigma$ values (error bars) in bins of 0.3~dex are displayed in green. The expected correlation for a constant abundance ratio of $X(N_2H^+)=7.5\times 10^{-10}$ is indicated by a dashed black line.}
	\label{fig:abundances}
\end{figure}

The good correlation found between the observed N$_2$H$^+$ intensities and total H$_2$ column densities derived in Fig.~\ref{fig:N2H+_cal} suggests that this molecule may only exhibit small abundance variations across the entire ISF region. In the absence of a direct calibration of the H$_2$ masses at the ALMA resolution, it is important to test the validity of this conclusion at the resolution of our single-dish observations.

In Figure~\ref{fig:abundances}, we display the estimated total N$_2$H$^+$ column densities detected in our single-dish maps ($N(N_2H^+)_{SD}$) in all positions shown in Fig.~\ref{fig:N2H+_cal} with at least one line component with S/N~$\ge$~3. These $N(N_2H^+)_{SD}$ values are derived following the approach presented by \citep[][see also Appendix~\ref{sec:fit_spectra:opacity}]{CAS02}.
If needed, the contribution of several lines are added in spectra with multiple components.
Each $N(N_2H^+)_{SD}$ measurement is compared with the corresponding total gas column density N(H$_2$) provided by previous {\it Herschel} measurements at the same position after the subtraction of the fit intercept (zero-level) derived in Eq.~\ref{eq:colden} (i.e. N(H$_2$)$_0=1.7\times10^{22}$~cm$^{-2}$).
For $log_{10}(N(N_2H^+))\sim[12.5,14.0]$ and $log_{10}(N(H_2)-N(H_2)_0)\sim[21.0,23.0]$ values, 
70\% of the respective column densities in OMC-2 are consistent, within a factor of two, with a constant abundance $X(N_2H^+)=7.5\times 10^{-10}$. The same relation is followed by the positions observed in OMC-1 despite their larger uncertainties in their molecular and dust emission properties.
A systematic deviation towards higher abundances is observed at $log_{10}(N(N_2H^+))> 14.0$ and $log_{10}(N(H_2)-N(H_2)_0)~\sim 23.0$, although the low number of points at these high column densities prevents any further analysis. Exceptions of this well-behaved correlation are also found at higher resolutions in the vicinity ($\lesssim$~20") of Orion BN/KL hot core with almost no N$_2$H$^+$ (1-0) emission in comparison with the detection of warm NH$_3$ gas \citep[e.g.,][]{GOD11}. With a little influence in our global analysis though, the local interpretation of our ALMA observations in this particular region should be then taken with caution.

According to our single-dish results,  N$_2$H$^+$ is found at relatively constant abundance at the typical gas column densities traced in the ISF (see also Fig.~\ref{fig:fits_opacities}). Moreover, its absolute value appears to vary less than 50\% compared to other filaments in clouds like Taurus \citep[with $X(N_2H^+)=5\times 10^{-10}$;][]{TAF15}. 
The above single-dish N$_2$H$^+$ abundances are also consistent with the results obtained from the comparison of SCUBA-derived H$_2$ column densities (assuming the same T$_{dust}$ values derived by \citet{LOM14}) and our new N$_2$H$^+$ ALMA observations. However, the different resolutions of both single-dish (with $\theta_{mb}$ between 36'' and 14'') and interferometric maps maps ($\theta_{mb}=$~4.5'') as well as the large uncertainties associated to these calculations (e.g., dust emissivity, filtering effects, etc...) prevent any further analysis.
Still, we remark here that the total H$_2$ masses in our maps are directly derived from the empirical relation between the observed N$_2$H$^+$ intensities and {\it Herschel} column densities defined by Eq.~\ref{eq:colden} avoiding the assumption of any particular N$_2$H$^+$ abundance. 
Although not directly involved in these latter estimates, the relatively stable N$_2$H$^+$ abundance in regions like OMC-1 and OMC-2 support the use of the integrated intensity of this molecule as a reliable proxy of the total gas column densities within the wide range of physical conditions considered in this study.

\end{appendix}

\end{document}